\documentclass{aa}
\usepackage{graphicx}
\usepackage{txfonts}
\usepackage{booktabs}
\usepackage{braket}
\usepackage[colorlinks=true, allcolors=blue]{hyperref}

\usepackage{graphicx}
\usepackage{pdfpages} 

\def\gsim{\mathrel{\spose{\lower 3pt\hbox{$\mathchar"218$}}
          \raise 2.0pt\hbox{$\mathchar"13E$}}}
\def\lsim{\mathrel{\spose{\lower 3pt\hbox{$\mathchar"218$}}
          \raise 2.0pt\hbox{$\mathchar"13C$}}}
          
\usepackage{verbatim,amscd, graphicx, amscd, mathrsfs}
\usepackage{amsmath}
\usepackage[english]{babel}
\usepackage[T1]{fontenc}
\usepackage{color}
\usepackage{listings}
\usepackage{lmodern}
\usepackage[colorlinks=true, allcolors=blue]{hyperref}
\usepackage{caption}
\usepackage{geometry}
\usepackage{graphicx}
\usepackage{multicol}
\usepackage{multirow}
\usepackage{float}
\usepackage{braket}
\usepackage{ulem}
\usepackage[utf8]{inputenc}

\newcommand{\Kd}{\texttt{Komrad dynamical}}
\newcommand{\Komrad}{\texttt{Komrad}}

\begin{document}

\title{Radiation-mediated shocks in gamma-ray bursts: Spectral evolution}

\author{Filip Alamaa$^{1,2,3}$ and Fr\'ed\'eric Daigne$^{1,4}$}

\authorrunning{F. Alamaa \& F. Daigne}
\titlerunning{Spectral evolution for RMS GRBs}

\institute{\textsuperscript{1}Sorbonne Universit\'e, CNRS, UMR 7095, Institut d'Astrophysique de Paris (IAP), 98 bis boulevard Arago, 75014 Paris, France\\
\textsuperscript{2}KTH Royal Institute of Technology, Department of Physics, SE-10691 Stockholm, Sweden\\
\textsuperscript{3}The Oskar Klein Centre for Cosmoparticle Physics, AlbaNova University Centre, SE-10691 Stockholm, Sweden\\
\textsuperscript{4}Institut Universitaire de France, Minist\`ere de l'Enseignement Sup\'erieur et de la Recherche, 1 rue Descartes, 75231 Paris Cedex F-05, France}

\abstract{Radiation-mediated shocks (RMSs) occurring below the photosphere in a gamma-ray burst (GRB) jet could play a crucial role in shaping the prompt emission. In this paper, we study the time-resolved signal expected from such early shocks. 
We model an internal collision using a 1D special relativistic hydrodynamical simulation, and we follow the photon distributions in the resulting forward and reverse shocks as well as in the common downstream region to well above the photosphere using a designated RMS simulation code. We compute the light curve and time-resolved spectrum of the resulting single pulse taking into account the emission at different optical depths and angles to the line of sight. For the specific case considered, we find a light curve consisting of a short pulse lasting $\sim 0.1~$s for an assumed redshift of $z = 1$, which could constitute a whole short GRB or be a building block within a highly variable longer GRB light curve. The efficiency is large, with $\approx 23$\% of the total burst energy being radiated. The spectrum has a complex shape at very early times, after which it settles into a more generic shape with a smooth curvature below the peak energy, $E_p$, and a clear high-energy power law that cuts off at $\sim 5~$MeV in the observer frame. The spectrum becomes narrower and softer at late times with $E_p$ steadily decreasing during the pulse decay from $E_p \approx 250~$keV to $E_p \approx 100~$keV. The low-energy index, $\alpha$, decreases during the bright part of the pulse from $\alpha \approx -0.5$ to $\alpha \approx -1$, although the low-energy part is better fit with a broken power law when the signal-to-noise ratio is high. The high-energy power law is generated by the reverse shock at low optical depths ($\tau < 30$) and has an index that decreases from $\beta \approx -2$ to $\beta \approx -2.4$. 
These results provide support for RMSs as potential candidates for the prompt emission in GRBs. 
}


\maketitle

\section{Introduction}

The emission mechanism responsible for the prompt phase in gamma-ray bursts (GRBs) has been a matter of discussion for many decades. The competing models can be divided into two families depending on the value of the optical depth at the emission radius towards a distant observer, $\tau$. In the optically thin family ($\tau < 1$), the models predict that the observed $\gamma$-rays are due to synchrotron radiation from high-energy electrons. The electrons can be accelerated either by internal shocks \citep{Narayan1992, ReesMeszaros1994, Kobayashi1997, DaigneMochkovitch1998} or by magnetic reconnection \citep{SpruitDaigneDrenkhahn2001, Drenkhahn2002, UhmZhang2014}. More recently, synchrotron emission from protons has also been discussed \citep{Ghisellini2020, Florou2021, Begue2022}.  
The optically thick family of models suggests that the observed radiation is emitted when the initially trapped photons decouple at the photosphere \citep[$\tau = 1$,][]{Paczynski1986, Goodman1986}. 

The typical observed GRB spectrum is much broader compared to thermal spectra \citep[e.g.][]{Yu2016}. The released emission must therefore be out of thermal equilibrium with the plasma for photospheric models to be viable. Since the radiation should be in thermodynamic equilibrium with the plasma close to the central engine, this requires that something disturbs the equilibrium close to the transparency radius \citep{ReesMeszaros2005}. Numerical simulations of GRB jets suggest that the jet is highly variable, which should lead to internal dissipation \citep{Lazzati2009, Lopez-Camara2013, Ito2015, Gottlieb2019}.

Shocks that occur below the photosphere in a GRB jet are mediated by radiation \citep{LevinsonBromberg2008}. The photon distribution downstream of a radiation-mediated shock (RMS) is very different compared to a synchrotron spectrum radiated downstream of an optically thin collisionless shock. Therefore, it is important to capture the correct shock physics if one wants to compare model predictions to data \citep[see][for a review on RMSs]{LevinsonNakar2020}. RMSs in GRBs have been studied extensively in the past \citep{Eichler1994, LevinsonBromberg2008, Budnik2010, Katz2010, Bromberg2011b, Levinson2012, KerenLevinson2014, Beloborodov2017, Ito2018, Lundman2018, Samuelsson2022, SamuelssonRyde2023, Wistemar2025b, Wistemar2025a}. However, the main focus in these studies has been the spectral shape rather than the time-dependent signal that one might expect from these events.

\begin{figure*}
\sidecaption
  \includegraphics[width=12cm]{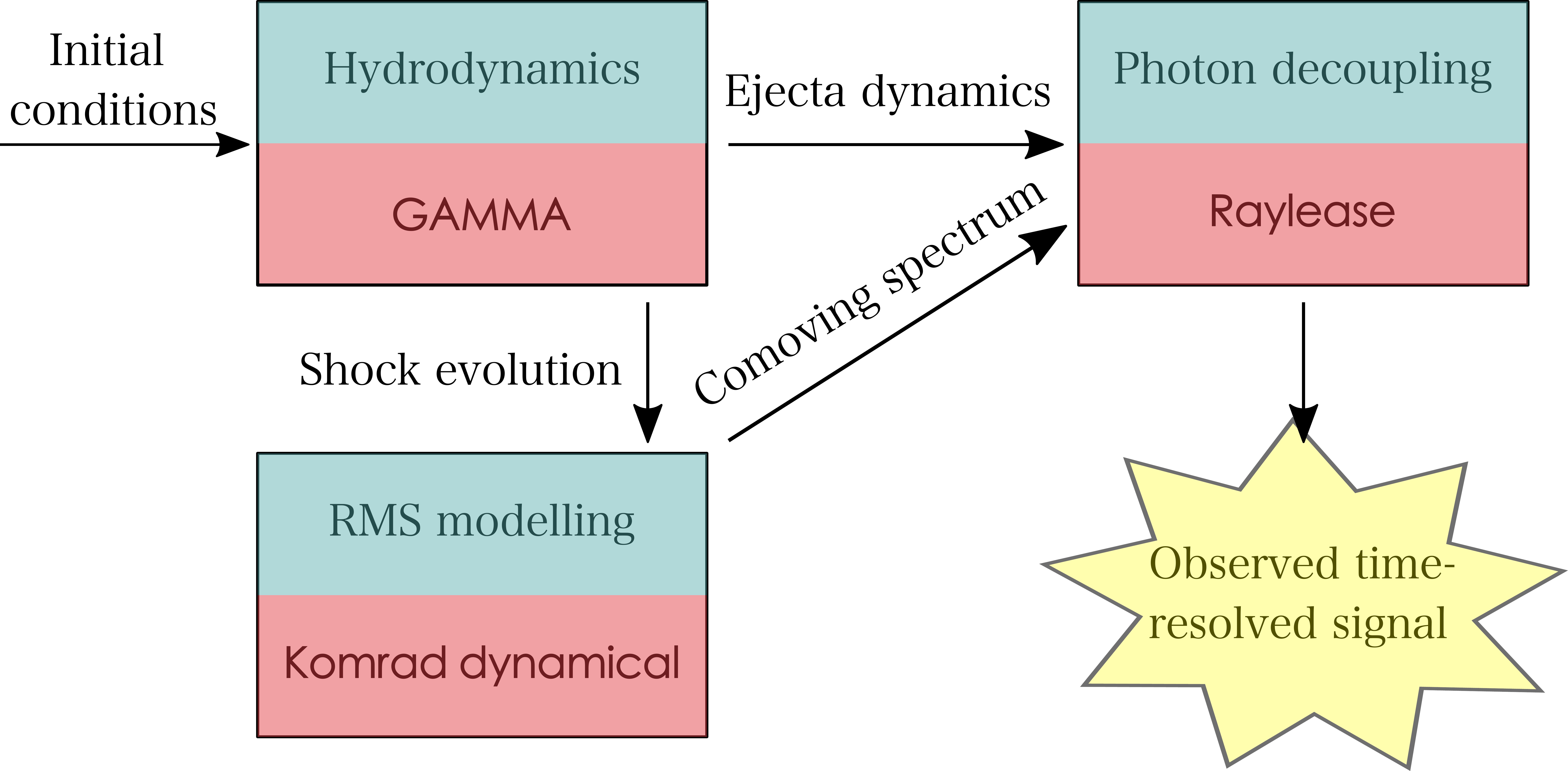}
     \caption{Flowchart showing the methodology of the paper. Each box shows one step of the chain, with the treated physics given in light blue and the simulation code used given in red. The final result is the time-resolved signal in the observer frame. The hydrodynamics are explained in Section~\ref{sec:hydro_implementation}, the RMS modelling in Section~\ref{sec:KRA_dynamical}, and the photon decoupling in Section~\ref{sec:time_resolved_signal}. The observed time-resolved signal is shown in Section~\ref{sec:results}.}
     \label{fig:flowchart}
\end{figure*}

In this paper, we study for the first time the time-resolved signal expected from a GRB photosphere including subphotospheric dissipation due to RMSs. To achieve this, we used a combination of different numerical codes, as outlined in the flowchart in Figure \ref{fig:flowchart}. 
First, we used a 1D special relativistic hydrodynamical code \citep[\texttt{GAMMA},][]{Ayache2022} to simulate an internal collision. 
The details regarding the hydrodynamical implementation are given in Section~\ref{sec:hydro_implementation}. The internal collision produces a reverse and a forward shock, and the two RMSs are modelled using the Kompaneets RMS approximation (KRA) developed in \citet{Samuelsson2022}. However, as explained in Appendix \ref{App:shock_broadening}, the radiation field in the shocks are not necessarily in steady state, and we have made several modifications and improvements to the KRA to tackle the problem at hand. The modifications, outlined in Section~\ref{sec:KRA_dynamical}, include the modelling of forward and reverse shocks, a dynamical (time-dependent) treatment of the two shocks, and a more realistic approach at low optical depths. The new version of the KRA is modelled using the code \Kd. 
The ejecta dynamics from the hydro simulation, together with the comoving photon distribution as a function of optical depth from the RMS modelling, are then used as input for a third code called \texttt{Raylease} \citep{Alamaa2024_intrapulse}. As explained in Section~\ref{sec:time_resolved_signal}, it calculates the observed time-resolved signal, accounting for a probabilistic photon decoupling as a function of both optical depth and angle. 

Our results are presented in Section~\ref{sec:results}. We find that the peak energy remains constant during the pulse rise time and starts to decrease at later times. At very early times, the spectrum has a complex spectral shape, with clear signatures from both the reverse and the forward shock. Later, the spectrum above the peak is well described by a power-law with an exponential cut-off, with both the spectral index and the cut-off energy decreasing as a function of time. The low-energy part consists of a smooth curvature, which may be approximated as a double power law. The results are discussed in Section~\ref{sec:discussion}, with emphasis placed on some key assumptions. Finally, we conclude in Section~\ref{sec:conclusion}. 

Temperatures in this paper denoted by $\theta$ are comoving and dimensionless and are given in units of electron rest mass energy as $\theta = k_{\rm B}T/m_ec^2$. Similarly, photon energies denoted by $\epsilon$ are comoving and dimensionless, defined as $\epsilon = h \nu /m_e c^2$. Energies denoted by $E$ are in the observer frame and given in units kiloelectronvolt. The outermost material in the ejecta is referred to as the front edge, and the innermost material, closest to the central engine, is referred to as the back edge.


\section{Hydrodynamical implementation}\label{sec:hydro_implementation}

\subsection{\texttt{GAMMA}}
The dynamics of the relativistic ejecta was computed in this paper using the publicly available code \texttt{GAMMA} \citep{Ayache2022}. \texttt{GAMMA} is a moving mesh special relativistic hydrodynamical code that is well suited for the treatment of GRB outflows. The code uses an arbitrary Lagrangian-Eulerian approach, which means that the mesh edges can be moved at any velocity during the dynamical evolution. Matching the velocity of the mesh edges to that of the flow makes the code pseudo-Lagrangian, which increases the numerical resolution \citep[][]{DuffellMacFadyen2011, DuffellMacFadyen2013, Duffell2016}. In 1D, the code is perfectly Lagrangian.

The code uses a finite-volume Gudonov scheme, and the fluxes at the cell interfaces are calculated using an approximate Riemann solver \citep[HLLC;][]{MignoneBodo2006}. 
\texttt{GAMMA} has been developed to follow the dynamics of internal shocks and of the jet deceleration by the ambient medium and to compute the associated radiation in the optically thin regime. It is well adapted to highly variable outflows \citep[see e.g.][]{Ayache2020}. Therefore, we used it here to get a handle on the GRB dynamics at lower radii, starting well below the photosphere.

A very important aspect when modelling the optically thick part of a GRB jet is to allow for the feedback of the radiation on the hydrodynamical properties. Unfortunately, \texttt{GAMMA} does not provide such feedback. This is a delicate point that is further discussed in Section~\ref{subsec:disc_radiation_feedback}. Therein, we argue that the feedback is not important in the current framework since we are not interested in exactly how the pressure, density, and velocity varies across the ejecta at any given time, but rather we care about the average values of these quantities in the two upstreams and the downstream. These average quantities should generally be accurate since they are determined by the continuity equations (shock-jump conditions).

\subsection{Initial conditions}\label{subsec:H_initial_conditions}
We considered a central engine that ejects relativistic material during a lab frame duration, $t_e$. The variability of the jet launching process at the central engine and the interaction of the jet and the surrounding material during the early propagation leads to variable observed injected power, ${\dot E}$, and observed injected mass flux, ${\dot M}$. 
As a consequence, the terminal Lorentz factor, $\eta \equiv {\dot E}/{\dot M}c^2$, also varies. 
The fluctuations of $\eta$ leads to shocks developing inside the jet, which reprocesses the kinetic energy of the outflow into internal energy, thereby shaping the observed emission. These internal shocks can occur both above and below the photosphere depending on the profiles of ${\dot E}(t)$ and ${\dot M}(t)$.

How ${\dot E}(t)$ and ${\dot M}(t)$ vary across the jet is not known, and their profiles are likely complex; see further discussion in Section \ref{subsec:disc_complicated_light_curve}. In this paper, we focused on one single internal collision leading to a single short pulse in the light curve. 
The mass flux was assumed constant over the ejecta duration. The injected energy flux was assumed to be a constant low value, ${\dot E}_s$, during an initial time, $t_s$, to smoothly increase during a transition time, $t_t$, and to remain constant at a high value, ${\dot E}_f$, during the remaining time, $t_f = t_e - (t_s + t_t)$. Specifically, the profile was given by
\begin{equation}\label{eq:init_Edot}
\begin{split}
	{\dot E}(t) &= {\dot E}_s, \quad {\rm if} \ 0 \leq t \leq t_s\\
	{\dot E}(t) &= \frac{{\dot E}_s + {\dot E}_f}{2} - \frac{{\dot E}_f - {\dot E}_s}{2} \\
    &\times \cos\left( \pi  \ \frac{t - t_s}{t_t} \right), \quad {\rm if} \ t_s < t < t_s + t_t\\
	{\dot E}(t) &= {\dot E}_f, \quad {\rm if} \  t_s + t_t \leq t \leq t_e.
\end{split}
\end{equation}
In the equation above, $t$ is the lab frame time that has elapsed since the front edge was emitted from the central engine. 
 
With the mass flux constant, the terminal Lorentz factor is directly proportional to ${\dot E}$. Specifying a minimum terminal Lorentz factor, $\eta_s$, fixes the constant mass flux as ${\dot M} = {\dot E}_s/\eta_s c^2$ and the maximum terminal Lorentz factor as $\eta_f = \eta_s {\dot E}_f/{\dot E}_s$. Since $\eta \propto {\dot E}$, Equation~\eqref{eq:init_Edot} implies a velocity stratification across the ejecta, which eventually leads to the formation of a reverse and a forward shock. We limited ourselves in the present study to the case where both shocks had crossed their respective masses before the ejecta reached the photosphere.

We assumed that the ejecta accelerates freely from an initial radius, $R_0$, where the Lorentz factor is $\Gamma_0$. The value of $R_0$ could potentially be comparable to a few times the Schwarzschild radius of the black hole \citep[e.g.][]{Iyyani2013}. However, numerical simulations suggest that the free acceleration is significantly delayed to $R_0 \sim 10^9-10^{10}~$cm due to high-pressure material surrounding the jet \citep{Gottlieb2019}. We assumed a pure fireball evolution, i.e. negligible magnetisation above $R_0$. At the start of the hydro simulation, the back edge had just reached the saturation radius, $R_s$, for the fast fireball as $R_{\rm min} = R_s (\eta_f) \equiv \eta_f R_0/\Gamma_0$. Thus, the simulation start time was $t_{\rm start} \approx R_{\rm min}/c + t_e$. 
The front edge was set as $R_{\rm max} = R_{\rm min} + ct_e$.\footnote{This prescription slightly overestimated the initial front edge radius, as it did not account for the actual velocity of the front edge between $R_0$ and $R_{\rm max}$. However, this slight overestimation does not affect the results.} 
The simulation used 3000 grid cells equally spaced in radius.

In each cell, the comoving density was set as \citep[e.g.][]{Piran1999}
\begin{equation}\label{eq:init_rho}
    \rho(R) = \frac{{\dot E}}{4 \pi R^2 c^3 \Gamma \eta},
\end{equation}
\noindent the comoving pressure as 
\begin{equation}\label{eq:init_p}
    p(R) = \frac{\rho c^2}{4} \left(\frac{R}{R_s}\right)^{-2/3} \left(\frac{\Gamma}{\eta}\right)^{-1/3},
\end{equation}
and the Lorentz factor as
\begin{equation}\label{eq:init_gamma}
    \Gamma(R) = \frac{\eta}{1+4\frac{p}{\rho c^2}}\, ,
\end{equation}
\noindent where $R_s \equiv \eta R_0/\Gamma_0$ is the saturation radius for a fireball with terminal Lorentz factor $\eta$.\footnote{Note that with this definition of $R_s$, the fireball still has slightly more internal than kinetic energy at $R_s$, with equipartition occurring at $R = \sqrt{2}R_s$.} 
These equations were obtained by assuming $\eta \gg 1$, a relativistic equation of state, and an adiabatic expansion of the fireball. 
Equations~(\ref{eq:init_rho}--\ref{eq:init_gamma}) result in a cubic equation for $(\Gamma/\eta)^{1/3}$, which close to the saturation radius has the real valued solution 
\begin{equation}\label{eq:cubic_solution}
    \frac{\Gamma}{\eta} = \frac{1}{3}\left(3 - C_1^2 + C_2 - \frac{6C_1^2 - C_1^4}{C_2}\right),
\end{equation}
where $C_1 \equiv R_s/R$ and 
\begin{equation}\label{eq:C2}
    C_2 \equiv \left(-C_1^6 + 9C_1^4 - \frac{27}{2}C_1^2 + \frac{3\sqrt{3}}{2}\sqrt{27C_1^4 - 4C_1^6}\right)^{1/3}.
\end{equation}
The above solution is valid when $C_1 < \sqrt{27/4}$ (corresponding to the regime where $4p/\rho c^2 < 3$). Equations \eqref{eq:init_rho}, \eqref{eq:init_p}, \eqref{eq:cubic_solution}, and \eqref{eq:C2}, together with $v = c\sqrt{1 - 1/\Gamma^2}$, determined the initial state in each cell.
Thus, the free parameters of the problem are $t_e$, $t_s$, $t_t$, ${\dot E}_s$, ${\dot E}_f$, $\eta_s$, $R_0$, and $\Gamma_0$.

\subsection{Boundary conditions}\label{subsec:boundary_conditions}
In this paper, we performed two simulations that differed only by their boundary conditions. The first simulation assumed that outside of the ejecta region on both sides, there existed a low density and low pressure material that propagated with a lower Lorentz factor. In practice, this was achieved by setting $p_{\rm ghost}/p_{\rm inside} = \rho_{\rm ghost}/\rho_{\rm inside} = 10^{-3}$ and $\Gamma_{\rm ghost} = \Gamma_{\rm inside}/2 + 0.55$. Here, the subscript ``ghost'' refers to properties in the ghost cells and the subscript ``inside'' refers to quantities in the grid cell closest to the ghost cell that is inside of the simulated region. The expression for $\Gamma_{\rm ghost}$ assured that it was always larger than unity. We used this expression instead of a vacuum at rest since it significantly increased the computational speed. These boundary conditions were supposed to mimic a vacuum outside of the ejecta region. This choice is motivated once the jet has escaped the surrounding high density material, originating from the parent star in the case of a collapsar progenitor or the tidally disrupted neutron star(s) in the case of a merger progenitor. 

The second simulation was motivated by numerical studies of collapsars, where it is commonly found that the jet at break out is highly variable \citep[e.g.][]{Lopez-Camara2014, Ito2015, Ito2019, Gottlieb2019, Gottlieb2020}. This variability would lead to multiple internal collisions occurring in the jet, with the single collision studied in this paper being one of these. 
This second simulation used a more appropriate periodic boundary condition,  where the properties of the ghost cells were determined by the properties in the cells on the other side of the simulated region. In practice, this was achieved by assuming that $\rho R^2$, $p R^{8/3}$, and $\Gamma$ were periodic. 

As discussed in Section \ref{subsec:disc_complicated_light_curve}, we found that the two simulations gave very similar results. Thus, we present only the results from the first simulation in Section \ref{sec:results}.

\subsection{Geometry of the ejecta}\label{subsec:H_geometry}

The hydrodynamical setup described in Section~\ref{subsec:H_initial_conditions} generated one reverse shock propagating to the left and one forward shock propagating to the right. Here, left and right are in terms of the Lagrangian mass coordinate, $m(R,t)$, defined as the total mass in the ejecta between the central engine and $R$ at time $t$ as
\begin{equation}
    m(R, t) = \int_0^R 4 \pi R^2 \Gamma \rho \, dR.
\end{equation}

One of the main purposes of the hydrodynamical simulation in this paper was to obtain the evolution of the shock properties as a function of radius, which served as input in the RMS modelling. In this regard, there were five different regions of interest in each snapshot of the hydrodynamical simulation. From left to right in mass coordinate, these were: i) the upstream of the reverse shock, ii) the immediate downstream of the reverse shock, iii) the far downstream region (common for both shocks), iv) the immediate downstream of the forward shock, and v) the upstream of the forward shock. Throughout the paper, hydrodynamical quantities are denoted by a subscript $u$ in the upstream regions i) and v), by a subscript $d$ in the immediate downstream regions ii) and iv), and by a subscript $\ast$ in the common far downstream region iii). A sketch of the Lorentz factor profile across the ejecta after both shocks have formed can be seen in the top panel in Figure~\ref{fig:KRA_geometry}, where the five regions have been marked.

\subsection{Output of the hydrodynamical simulation}\label{subsec:H_hydro_outputs}
This subsection lists the hydrodynamical properties that are of interest for the rest of the paper. 
To model the evolution of the comoving photon distribution in \Kd, we needed information regarding the structure of the two shocks, the scattering timescale, and the overall cooling of the ejecta due to adiabatic expansion.
The latter was quantified using the parameter $\varphi$, defined as
\begin{equation}\label{eq:varphi}
    \varphi \equiv - \frac{1}{3}\frac{d\log \rho}{d\log R}.
\end{equation}
\noindent In an idealised outflow where the comoving density decreases as $\rho \propto R^{-2}$, one gets $\varphi = 2/3$.

The scattering timescale, $t_{\rm sc}$, is important since it determines the time it takes for the radiation in each shock to reach steady state, as well as the level of thermalisation experienced by the downstream radiation field. In the observer frame, it is given by
\begin{equation}\label{eq:t_sc}
    t_{\rm sc} = \frac{\Gamma}{\kappa \rho c},
\end{equation}
\noindent where $\kappa$ is the opacity, and we assumed $\Gamma \gg 1$. In this paper, we use $\kappa = 0.4~$cm$^2/$g.

When the scattering time is much shorter than the dynamical time, the radiation in the RMS is in steady state as explained in Appendix~\ref{App:shock_broadening}. Each steady-state RMS is characterised by three parameters, which can be taken to be the dimensionless upstream velocity as measured in the shock rest frame, $\beta_u$, the comoving upstream temperature, $\theta_u$, and the photon-to-proton ratio, $\zeta$, \citep[e.g.][]{Lundman2018}. The exact calculation of these three quantities from the hydrodynamical output is detailed in Appendix~\ref{App:hydro_to_RMS}. In short, the upstream velocity was obtained from the shock jump conditions since the increase in pressure and density across the shock was known. The upstream temperature was given by the ratio of the upstream pressure to the upstream particle number density. The photon-to-proton ratio at mass coordinate $m$ was obtained by assuming a thermodynamic equilibrium at $R_0$, after which $\zeta$ was calculated as a function of ${\dot E}$ and $\eta$. The three RMS parameters were at any given time different for the two shocks, and they varied as a function of radius for both shocks.

To obtain the time-resolved signal in the observer frame in Section~\ref{sec:time_resolved_signal}, we needed to compute the evolution of the optical depth within the jet. 
Commonly, the radial optical depth from $R$ towards a distant observer is estimated as $\tau = \kappa \rho R/\Gamma$ \citep[e.g.][]{Beloborodov2011}.
However, for the vacuum boundary case considered in Section \ref{sec:results}, the ejecta was very thin, making photon escape at the front edge important. Thus, we calculated $\tau$ as \citep{Abramowicz1991, DaigneMochkovitch2002}\footnote{The definition in Equation~\eqref{eq:optical_depth} differs by a factor of two from that used in \citet{Abramowicz1991, DaigneMochkovitch2002}. This was to be consistent with the expression of $\tau$ used in the decoupling probability distribution $f(\tau, \mu^\prime)$ in Equation \eqref{eq:observed_flux} \citep[see eq. (B6) in][]{Beloborodov2011}.}
\begin{equation}\label{eq:optical_depth}
    \tau(R, t) = \int_R^{R_{\rm esc}} \frac{\kappa\rho}{\Gamma} \, dR,
\end{equation}
\noindent where $R_{\rm esc}$ is the radius where a radially moving photon injected at $R$ and $t$ reaches the front edge of the ejecta.

With the definition in Equation~\eqref{eq:optical_depth}, each mass coordinate had one associated value for the optical depth at each time. However, to obtain the observed signal, we wished to associate the whole ejecta at each lab frame time with a single value of the optical depth. This value was taken to be the mass averaged value of the optical depth. This was an approximation since the optical depth actually varied significantly across the ejecta at any given time. Specifically, the front edge had a lower optical depth compared to the rest of the ejecta, which meant that in reality one might expect a thermal precursor to the main signal. Considering such effects are unfortunately outside the scope of the current paper. 

Using the mass averaged value, the optical depth became a function of the lab frame time only, $\tau(t)$. Thus, we could also obtain the lab frame time when the ejecta had reached a specific optical depth, $t(\tau)$. This quantity is used in the Section \ref{sec:time_resolved_signal}.
Each snapshot of the hydro also needed to be associated with a single radius, which was taken to be the mass averaged value over the ejecta. This was valid since the width of the ejecta, $\Delta R$, was at all times much smaller than the current radius, $\Delta R/R \ll 1$.

\section{The Kompaneets RMS approximation in a dynamical flow}\label{sec:KRA_dynamical}

The hydrodynamical simulation described in the previous section gave us a handle on the dynamics of the GRB ejecta. 
As we considered initial conditions leading to the formation of shocks at high optical depth, we needed to model the evolution of the photon distribution below, at, and above the photosphere to get accurate predictions for the observed time-resolved signal.
Crucially, we needed to capture how the photon distribution changes due to the subphotospheric reverse and forward shocks, where photons can be repeatedly scattered in a steep velocity gradient, significantly altering their energies. 

The Kompaneets RMS approximation \citep[KRA,][]{Samuelsson2022} is an efficient way to model radiation in this context, without the need for computationally expensive simulations. The KRA was developed in \citet{Samuelsson2022} assuming a steady jet (constant $\Gamma$ and $\rho \propto R^{-2}$). Furthermore, the shock was assumed to dissipate all of its energy during one doubling of the radius. Therefore, the shock life-time was comparable to the expansion time of the jet, and potential dynamical effects on the shock itself were ignored. 
However, as derived in Appendix~\ref{App:shock_broadening}, the radiation in the shock goes from being in a steady state to being dynamically evolving already at an optical depth of $\tau = 1/\beta_u^2$ for a steady jet. Thus, this transition can happen far below the photosphere in non-relativistic RMSs.
To account for this, as well as the fact that the ejecta properties evolved significantly, i.e. the jet was not steady, we developed \Kd, which is a numerical code that utilises the KRA to simulate RMSs in a dynamical environment. All corresponding modification are detailed in Appendix~\ref{app:KRA_dynamical}.
In the following, we start by giving a brief description of the KRA (the interested reader is referred to \citet{Samuelsson2022} for more details), after which the most significant modifications are summarised.

\subsection{A brief summary of the KRA}\label{subsec:KRA_breif_summary}

When photons traverse an RMS, they Compton scatter in the velocity gradient across the shock. This is a Fermi-like process in which the photons gain energy on average with each scattering. For non-relativistic and mildly relativistic RMSs, the relative photon energy gain in a scattering is independent of the pre-scattered photon energy. 
This leads to the development of a power-law photon distribution in the shock region, extending between two characteristic energies, $\epsilon_{\rm min}^{\rm RMS}$ and $\epsilon_{\rm max}^{\rm RMS}$ \citep{Ito2018, Lundman2018, Samuelsson2022}. 

The above process is very similar to thermal Comptonisation of photons with hot electrons, which is also characterised by Compton scattering and an energy-independent average increase of the photon energy per scattering. This is the idea behind the KRA. The KRA models thermal Comptonisation using the Kompaneets equation, which is numerically easy to solve. The RMS transition region is split into three zones, the upstream zone, the RMS zone, and the downstream zone. Thermal photons are injected from the upstream zone into the RMS zone. In the RMS zone, the photons interact with hot electrons, which increases the average photon energy. The energy gain in the RMS zone mimics the energy gain a photon would experience due to bulk Comptonisation in a real RMS. Photons have a non-zero probability per scattering time to be advected from the RMS zone into the downstream zone. Since the escape probability is energy independent, a power-law spectrum develops in the RMS zone between two characteristic energies, $\epsilon_{\rm min}^{\rm KRA}$ and $\epsilon_{\rm max}^{\rm KRA}$.

The steady state RMS zone spectrum in the KRA is characterised by three parameters: the comoving upstream zone temperature $\theta_{u,{\rm K}}$,\footnote{It is necessary to add the subscript K to the upstream temperature $\theta_{u,{\rm K}}$, since it is in fact slightly higher compared to the upstream temperature obtained from the hydro $\theta_u$, see Equation~\eqref{eq:theta_u_K}.} the effective temperature in the RMS zone, $\theta_r$, and the Compton $y$-parameter of the RMS zone, $y_r$. Here, the subscript $u$ when used for KRA quantities refers to the upstream zone, corresponding to regions i) and v) in the hydro notation, and the subscript $r$ refers to the RMS zone, which is situated in between the upstream and the immediate downstream regions, i.e. between regions i) and ii), and iv) and v) in the hydro notation. The subscript $\ast$ is used when referring to quantities in the downstream zone, which corresponds to region iii) in the hydro notation. 

The photons in the upstream zone are assumed to be in a thermal Wien distribution at temperature $\theta_{u,{\rm K}}$. Since the average photon energy in a Wien distribution equals three times the temperature, the lower characteristic energy is given by $\epsilon_{\rm min}^{\rm KRA} = 3\theta_{u,{\rm K}}$. The maximum energy in the RMS zone is reached when the energy gain in a scattering event is balanced by the energy loss due to electron recoil. This gives $\epsilon_{\rm max}^{\rm KRA} = 4\theta_r$ \citep{RybickiLightman1979, Samuelsson2022}. The parameter $y_r$ is a measure of the average photon energy increase within the RMS zone. With $\epsilon^{\rm KRA}_{\rm min}$ and $\epsilon^{\rm KRA}_{\rm max}$ set by the other two parameters, the effect of $y_r$ is to regulate the slope of the power-law segment in between the two characteristic energies. 

To get the KRA to produce similar spectra to an RMS, one requires $\epsilon_{\rm min}^{\rm KRA} = \epsilon_{\rm min}^{\rm RMS}$, $\epsilon_{\rm max}^{\rm KRA} = \epsilon_{\rm max}^{\rm RMS}$ and 
that the average energy of the radiation advected into the downstream is similar in both frameworks. Via analytical arguments and an empirical study, an accurate conversion between the KRA parameters and the physical RMS parameters was found in \citet{Samuelsson2022} (for completeness, it is also given in Appendix \ref{subsec:H_KRA_RMS_to_KRA_conversion}). With this conversion, the steady-state output of the KRA agreed very well with the output from a special relativistic, Lagrangian radiation hydrodynamical code \citep[\texttt{radshock},][]{Lundman2018} \citep[see Figures 2 and 3 in][]{Samuelsson2022}. However, the simulation time was drastically decreased by as much as 4--5 orders of magnitude, highlighting the strength of the KRA. The parameter correspondence is one-to-one, which means that each steady state RMS spectrum corresponds to one unique set of KRA parameters.

The KRA is applicable to RMSs in weakly magnetised environments, where the incoming upstream plasma velocity satisfies $\gamma_u \beta_u \lesssim 3.5$ as measured in the stationary shock frame.\footnote{Although the jet is ultra-relativistic, shocks due to internal motion in the jet are unlikely to be more than mildly relativistic as measured in the shock frame \citep[e.g.][]{Samuelsson2022}.} As expected for RMSs in GRBs, the shock is assumed photon rich, which means that the main source of photons is advection of existing upstream photons rather than photon generation inside the shock and the immediate downstream \citep{Bromberg2011b}. 

\subsection{Major modifications in KRA dynamical}\label{subsec:KRA_modifications}
In this subsection, we summarise some of the most significant modifications to the KRA in a dynamical flow. All of the details are given in Appendix~\ref{app:KRA_dynamical}.

\begin{figure}
\begin{centering}
	\includegraphics[width=\columnwidth]{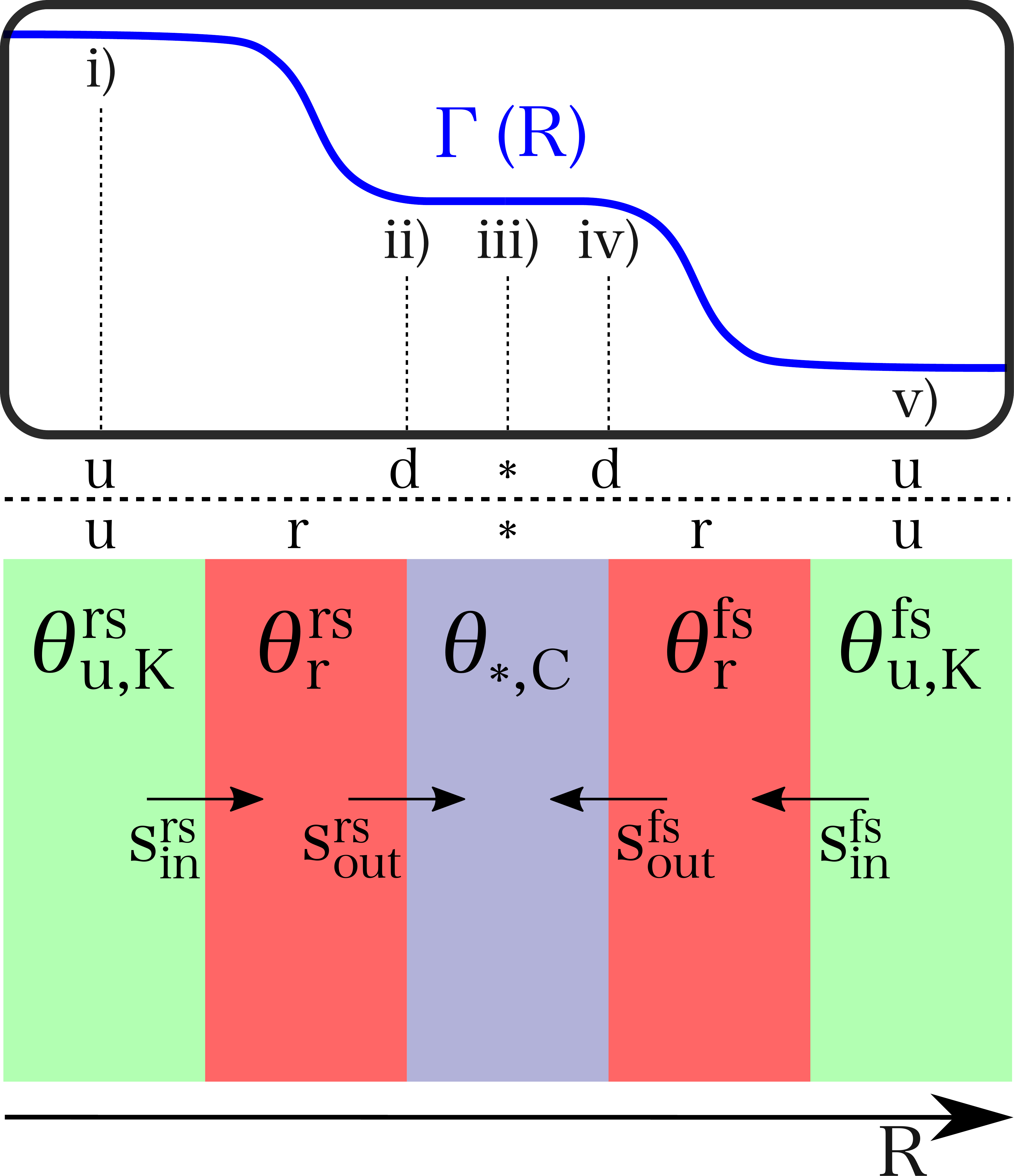}\\
\caption{Top panel: Sketch of the Lorentz factor profile across the ejecta with the positions of the five different regions mentioned in subsection \ref{subsec:H_geometry} marked. The dashed vertical lines indicate the position of each region relative to the KRA zones given in the bottom panel. Bottom panel: Geometry of the KRA with both reverse and forward shocks included. Green, red, and purple colours indicate the upstream zones, the RMS zones, and the common downstream zone, respectively. The zones are coupled via source terms as indicated in the figure. The RMS and downstream zones are evolved using the Kompaneets equation (Equation~\eqref{eq:Kompaneets_sph}) while the upstream zones are assumed to be in a thermal Wien distribution. The relevant temperature in each zone is given. This panel is adapted from \citet{Samuelsson2022}, which shows the geometry of the KRA with three zones. In between the two panels, the subscripts used to refer to the different regions and zones are shown for clarity.
}
    \label{fig:KRA_geometry}
\end{centering}
\end{figure}

The KRA has been updated such that it can model two shocks if both a reverse and a forward shock are present. In case of two shocks, the KRA consists of five different zones that are coupled via source terms. Radiation flows from the upstream zones, via the two shock zones, to the common downstream zone. The common downstream is treated as a single fluid with one collective electron temperature. This temperature is not a free parameter. Rather, the temperature is assumed to equal the downstream Compton temperature, $\theta_{\ast, {\rm C}}$, defined as the temperature the electrons get in Compton equilibrium with the current downstream radiation field. This is a good approximation since the electrons are tightly coupled to the radiation and their thermalisation timescale is incredibly short compared to the photon scattering time \citep{Lundman2018}. A schematic of the geometry with the five zones, and their positions relative to the hydrodynamical ejecta, is given in Figure~\ref{fig:KRA_geometry}.

The previous version of the KRA assumed that the jet was steady and that the upstream properties did not vary significantly over the shock life-time. In practice, this meant that $\beta_u$ and $\zeta$ were both constant, and $\varphi = 2/3$ such that all photon energies and the upstream temperature decreased as $\epsilon, \theta_{u,{\rm K}} \propto R^{-2/3}$. In the modified version, $\beta_u$, $\zeta$, $\theta_u$, and $\varphi$ can all vary with radius. In the current paper, the evolution of each parameter was given by the hydrodynamical simulation, as explain in Section~\ref{subsec:H_hydro_outputs}.

When an RMS reaches lower optical depths, it broadens \citep{Levinson2012}. An analytical description of this is given in Appendix~\ref{App:shock_broadening} but one can get an intuitive idea for the origin of this broadening from the following argument. For an RMS in steady state, there is a necessary number of scatterings that the photons have to make in order to mediate the shock and stop the incoming flow (as seen in the shock stationary frame). As the scattering length increases at lower optical depths when the plasma density decreases, the shock transition region must increase. The dynamical version of the KRA accounts for this broadening by varying the source terms as a function of radius such that the RMS zone always contains the appropriate number of photons. 

The upstream zone contains a finite amount of photons, $N_{u, {\rm tot}}$. When this zone is depleted, the RMS dissolves. In \Komrad \ (the name of the original simulation code), the photons in the RMS zone were instantaneously removed following the upstream depletion. However, if this happens at low optical depths, the shocks can contain a large number of photons due to the aforementioned RMS broadening \citep{Levinson2012}. This progressive dissolution of the RMS is taken into account in \Kd: the photon distribution in the RMS zone continues to evolve even when the upstream zone is depleted, while the remaining photons are slowly advected from the RMS into the downstream zone.

\subsection{Input and output}\label{subsec:KRA_input_output}
The input parameters required to run \Kd \ is $\theta_u$ and $N_{u, {\rm tot}}$ for the upstream zones, $\beta_u$ and $\zeta$ for the RMS zones, and $\varphi$ and $t_{\rm sc}$ for both the RMS zones and the downstream zone. All of these parameters were obtained from the hydro simulation as explained in Appendix~\ref{App:hydro_output_to_KRA_input}.

The output of \Kd \ is the photon spectral number density as a function of radius, $N_\epsilon(R)$, in the five different zones (see Appendix~\ref{app:KRA_dynamical} for details). 
The photon spectral number density, together with the information regarding the dynamics from \texttt{GAMMA}, such as $\tau(R)$ and $\Gamma(R)$, was used to obtain the time-resolved signal as explained in the next section.

\begin{table*}
    \centering
    \caption{Parameters used in this paper. The values result in a terminal Lorentz factor in the fast material of $\eta_f = 400$.} 
    \begin{tabular}{ccccccccc}
    \hline
    \hline
      $t_e$ [s]& $t_s$ [s] & $t_t$ [s] & ${\dot E}_s $ [erg/s] & ${\dot E}_f$ [erg/s] & $\eta_s$ & $R_0$ [cm] & $\Gamma_0$ & $z$\\
    \hline
      $0.15$
    & $0.075$
    & $0.045$
    & $10^{52}$ 
    & $10^{53}$ 
    & $40$
    & $3\times10^9$ 
    & $4$
    & $1$ \\
    \hline
    \end{tabular}
    \label{tab:params}
\end{table*}
\begin{figure}
    \centering
    \includegraphics[width = 0.99\columnwidth]{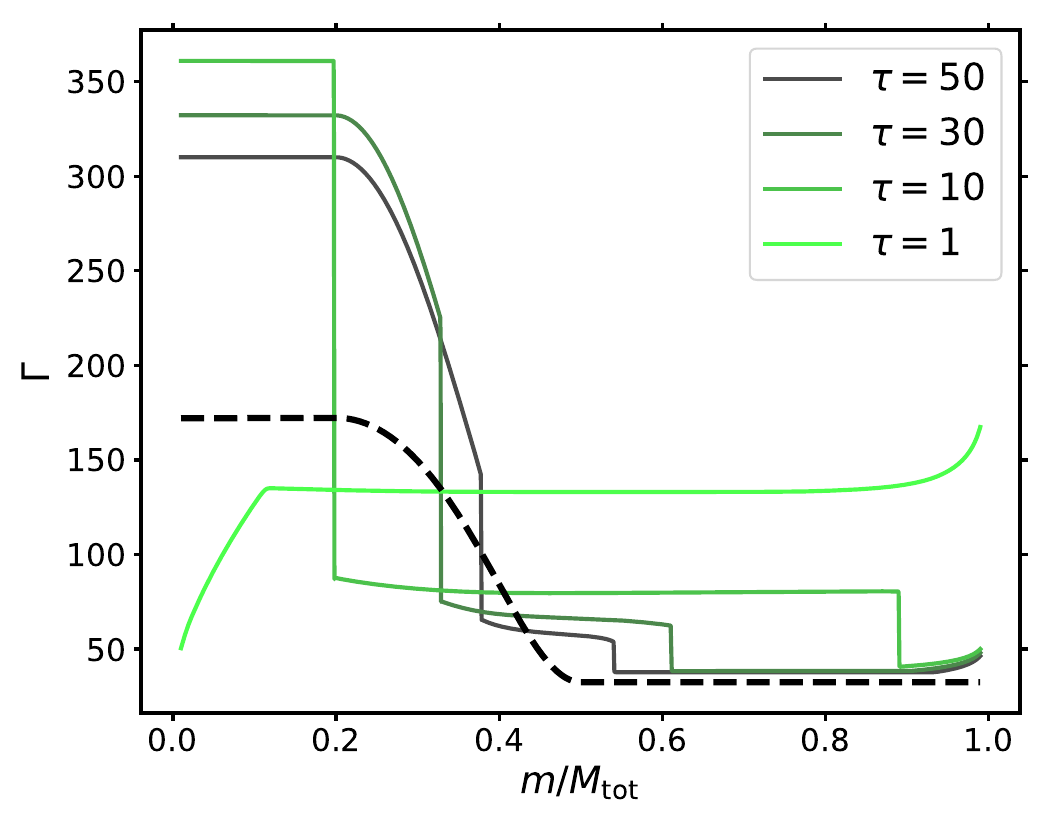}
    \includegraphics[width = 0.99\columnwidth]{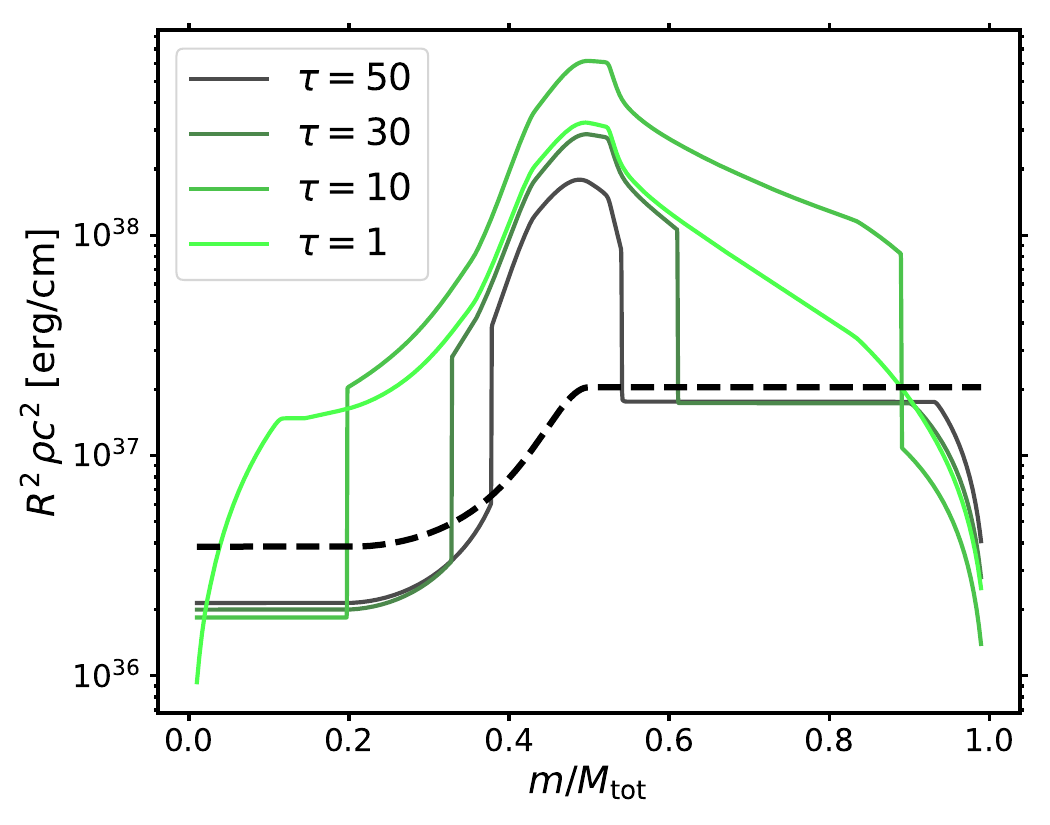}\\
    \caption{Lorentz factor (top) and the comoving density (bottom) at different optical depths as obtained from the hydrodynamical simulation using the vacuum boundary conditions. The dashed black line gives the corresponding profile at the start of the simulation. The comoving photon distributions in the five different zones of \Kd \ at the corresponding times are given in Figure~\ref{fig:KRA_comoving_photon_distribution}. 
    The simulation parameters are given in Table~\ref{tab:params}.}
    \label{fig:hydro_evolution}
\end{figure}

\section{Observed time-resolved signal}\label{sec:time_resolved_signal}

Photons make their last scattering at different optical depths and different angles compared to the line-of-sight due to the probabilistic nature of the decoupling process. Thus, the signal at any given observer time is a superposition of comoving spectra at different optical depths and different Doppler boosts \citep{Peer2008, Beloborodov2011}. The first photons to reach the observer are those that decouple at high optical depths and at the line-of-sight. These are followed by photons emitted at progressively lower optical depths (larger radii) and larger angles \citep{PeerRyde2011}. If the comoving photon field evolves significantly across the region where photons make their last scattering, the observed signal will evolve as a function of time \citep{Alamaa2024_intrapulse}. 

The photon spectral flux, $F_\epsilon(t_{\rm obs})$, at observer time $t_{\rm obs}$ was obtained by integrating the comoving photon distribution over all optical depths and angles, accounting for the probability of a photon making its last scattering at a given optical depth and angle, as well as the predicted arrival time as \citep[see e.g. eq. (A5) in][]{Alamaa2024_intrapulse}
\begin{equation}\label{eq:observed_flux}
\begin{split}
    F_\epsilon(t_{\rm obs}) &= \frac{1+z}{4 \pi d_l^2} \int_0^\infty d\tau \int_{-1}^1 d\mu^\prime \int_{R_{\rm back}}^{R_{\rm front}} dR \\
    & \frac{f(\tau, \mu^\prime)}{\Delta R} \epsilon^{\prime} N^\prime_{{\rm tot},\epsilon} \, \delta\left(\frac{t_{\rm obs}}{1+z} -t_\delta \right).
\end{split}
\end{equation}
\noindent In the equation above, $z$ is the redshift, $d_l$ is the luminosity distance, $\mu^\prime = \cos(\theta^\prime)$, with $\theta^\prime$ being the comoving angle between the radial direction and the line-of-sight, and $f(\tau, \mu^\prime)d\tau d\mu^\prime$ is the probability for a photon to make its last scattering between optical depth $\tau$ and $\tau + d\tau$ and angle $\mu^\prime$ and $\mu^\prime + d\mu^\prime$. The probability distribution $f(\tau, \mu^\prime)$ was derived in \citet{Beloborodov2011} and rewritten as a function of optical depth in \citet{SamuelssonRyde2023}.\footnote{The calculation of the probability distribution in \citet{Beloborodov2011} assumed a steady wind (constant $\Gamma$ and $\rho \propto R^{-2}$), which is not the case in the current paper. However, given the definition of optical depth, the probability distribution $f(\tau, \mu^\prime)$ should not be largely affected by these differences and here we used the expression unchanged.} 
$R_{\rm front}$ and $R_{\rm back}$ are the radial position of the front and back edge of the ejecta, respectively, both evaluated at time $t(\tau)$, and $\Delta R = R_{\rm front} - R_{\rm back}$ is the instantaneous ejecta width. 
The photon spectral number density $N^\prime_{{\rm tot}, \epsilon}$ is that of the whole ejecta (the sum of the photon number densities in all five KRA zones) and it is evaluated at an energy $\epsilon^\prime = (1+z)\epsilon/{\mathcal D}$, where ${\mathcal D} = \Gamma(1+\beta\mu^\prime)$ is the Doppler factor.
The time $t_\delta$ in the Dirac delta function is given by
\begin{equation}
    t_\delta = t(\tau) - \frac{R}{c}\left[ \frac{\mu^\prime +\beta}{1 + \beta \mu^\prime}\right].
\end{equation}

To solve Equation~\eqref{eq:observed_flux}, we used the code \texttt{Raylease} developed in \citet{Alamaa2024_intrapulse} (see Appendix A therein). The parameters $\Delta R$, ${\mathcal D}$, $\beta$, and $t(\tau)$ were readily taken from the hydrodynamical simulation. The Lorentz factor $\Gamma$ and velocity $\beta$ used were those for the downstream region, which contained essentially all of the radiation at the relevant optical depths where the photons escaped. The comoving photon distribution $N_{{\rm tot},\epsilon}$ was obtained from the KRA.

\section{Results}\label{sec:results}
\begin{figure*}
    \centering
    \includegraphics[width = 0.99\columnwidth]{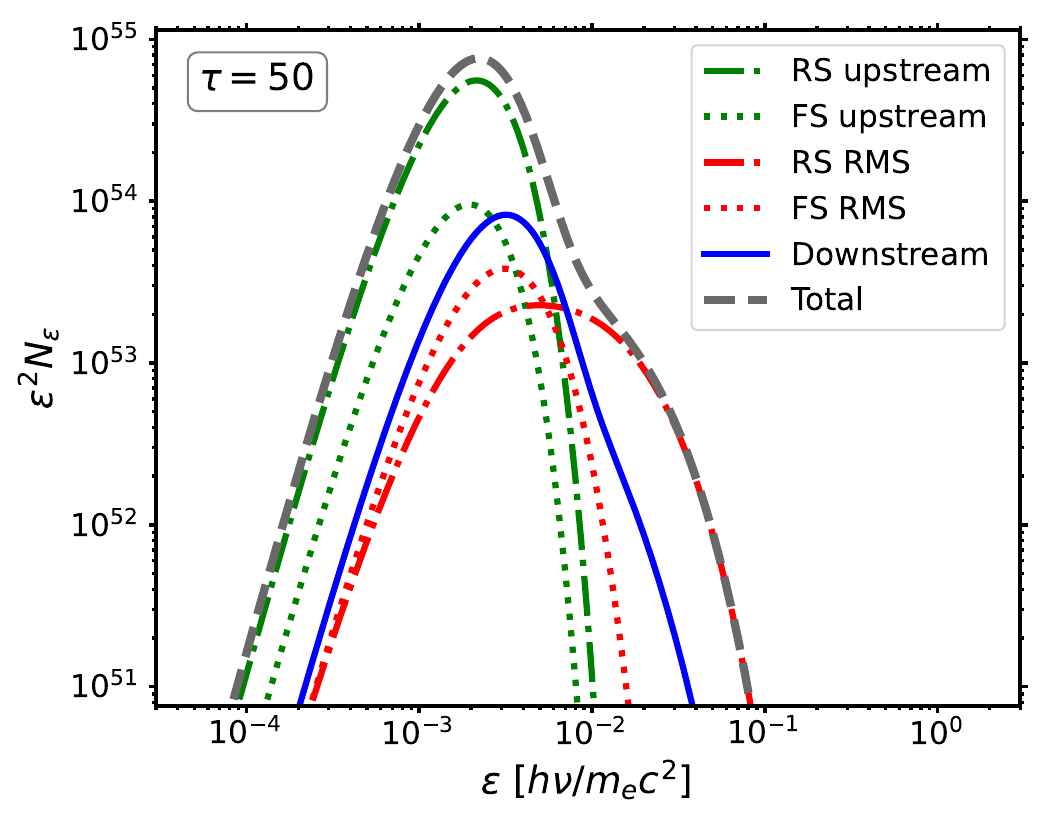}
    \includegraphics[width = 0.99\columnwidth]{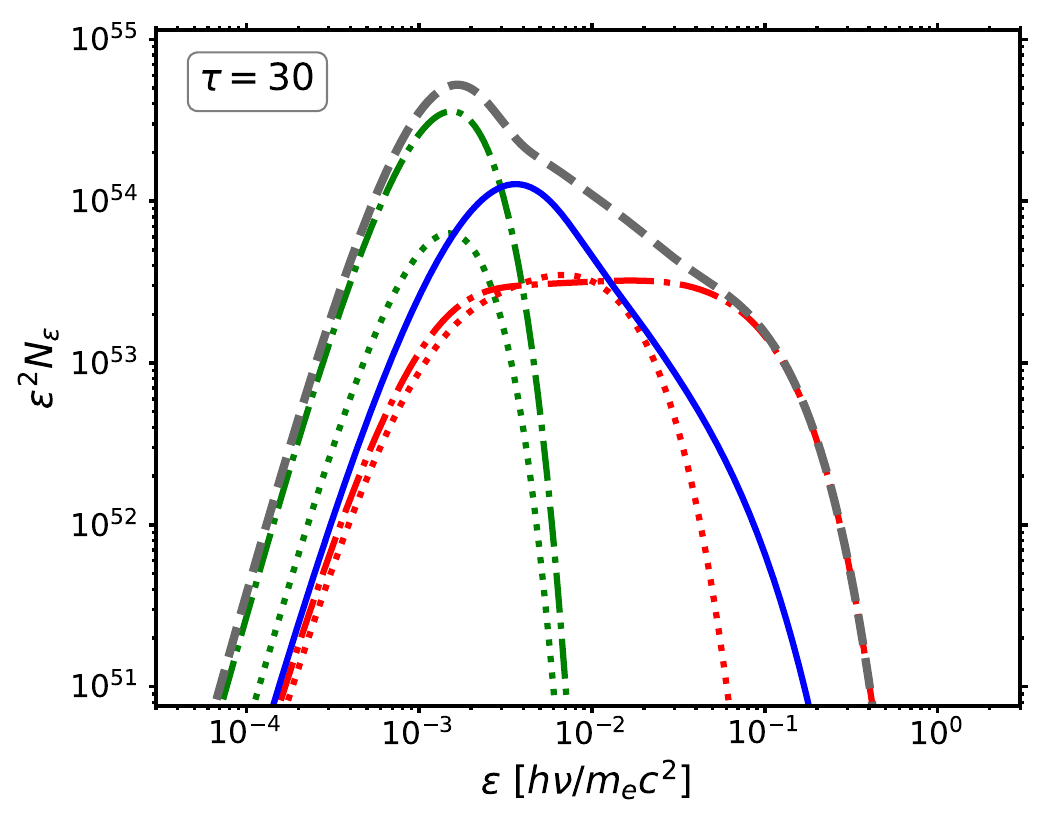}\\
    \includegraphics[width = 0.99\columnwidth]{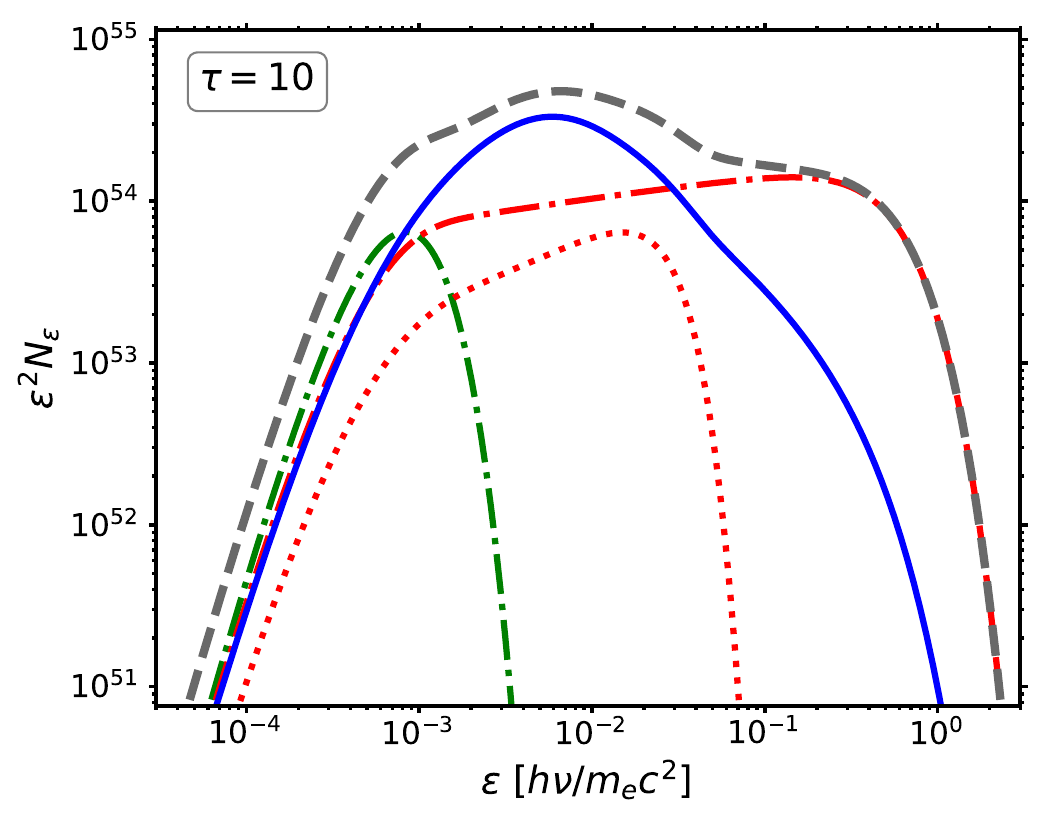}
    \includegraphics[width = 0.99\columnwidth]{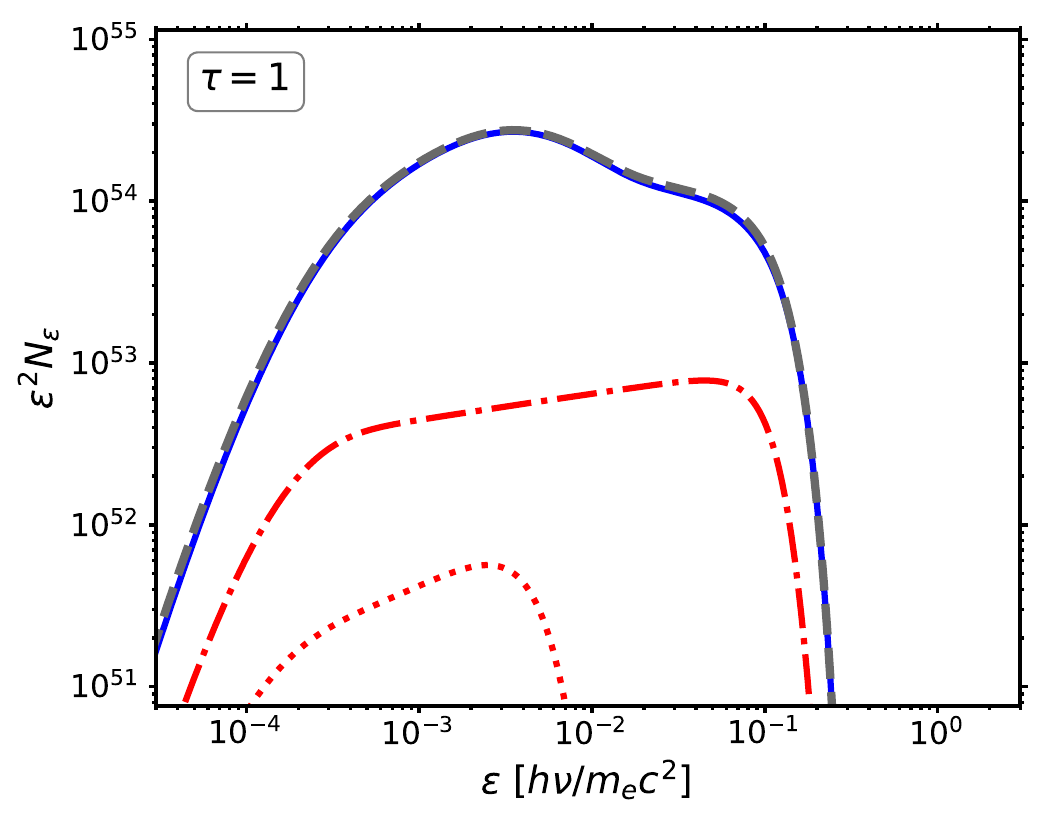}\\
    \caption{Comoving photon distribution in the five different zones as obtained by \Kd. The line coding is shown by the legend in the top-left panel, where RS and FS stand for reverse shock and forward shock, respectively. The four different panels show the distributions at different optical depths, corresponding to the optical depths in Figure~\ref{fig:hydro_evolution}. The observed photon distribution consists of a superposition of comoving spectra, leading to a smoothening and broadening effect on the observed spectrum. The simulation parameters are given in Table~\ref{tab:params}.}
    \label{fig:KRA_comoving_photon_distribution}
\end{figure*}

The results presented in this section are for the simulation using the vacuum boundary conditions, as explained in Section \ref{subsec:boundary_conditions}. The results from the simulation using the periodic boundary conditions were very similar.

\subsection{Ejecta evolution}
In Figures~\ref{fig:hydro_evolution} and \ref{fig:KRA_comoving_photon_distribution}, we show the evolution of the physical conditions in the ejecta. The parameters for the run are given in Table~\ref{tab:params}. Figure~\ref{fig:hydro_evolution} shows the Lorentz factor (top) and density (bottom) at different optical depths. The distributions at the start of the simulation ($\tau \approx 430$) are given by the black dashed lines. At the start of the simulation the ejecta was still accelerating. Notably, the initial Lorentz factor in the fast part of the ejecta was less then half of its terminal value, $\eta_f = 400$. 
The acceleration was gradual and the terminal value was never quite reached due to presence of the shocks. A maximum value of $\Gamma \simeq 370$ was reached at the reverse shock crossing. 
The forward shock was weaker and it reached the edge of the ejecta first. Once the forward shock had crossed, the downstream region accelerated freely and the average Lorentz factor increased at the expense of the ejecta internal energy.

The density ($\rho R^2$) initially increased drastically in the intermediate region between the slow and the fast part due to adiabatic compression. The compression led to an increase in temperature and pressure, and the shocks formed once the immediate downstream pressure had become comparable to the incoming ram pressure. As the left-hand side started off hotter than the right-hand side, the reverse shock formed slightly earlier than the forward shock. The density continued to increase after the shocks had been launched albeit more slowly. When the forward shock had crossed, the acceleration of the downstream led to a decrease in the density, which in turn led to a quick drop in the optical depth. This means that there can be a tendency for RMSs to finish their dissipation close to the photosphere, see further discussion in Section~\ref{subsec:disc_high_efficiency}.

Figure~\ref{fig:KRA_comoving_photon_distribution} shows the comoving photon distribution in the five zones at the corresponding optical depths, as obtained by \Kd. In the first snapshot, most of the photons were situated in the two upstreams zones. Bulk Comptonisation in the two shocks had just started and the photon distributions in the RMSs were still quasi-thermal. At optical depth $\tau = 30$, the reverse and forward shocks had developed further, leading to a non-thermal shape. There were still a significant fraction of photons in the two upstreams zones, leading to a thermal bump around $\epsilon \gtrsim 10^{-3}$. The downstream spectrum had begun to broaden but was still quite narrow. 

At $\tau = 10$, the reverse shock had developed a power-law from bulk Comptonisation that stretched more than two orders of magnitude in energy. Its upper cut-off energy continued to increase, which was due to two reasons: i) as the density decreased, the shock front sped up \citep{Sakurai1960}, which increased the maximum photon energy in the shock \citep[e.g.][]{Samuelsson2022}. ii) The photons had a progressively harder time to mediate the shock when the ejecta approached the photosphere. As the scattering frequency decreases, the relative energy transfer per scattering must increase for the radiation to decelerate the incoming flow. Since the maximum energy is proportional to the relative energy gain per scattering \citep{RybickiLightman1979, Samuelsson2022}, this further increased the upper cut-off energy in the RMS. This latter behaviour is apparent from Equation~\eqref{eq:deps_d_dr} and became important when the RMS entered the dynamical stage.

That the maximum energy in the RMS increases at lower optical depths is in direct contrast to the behaviour of a photon poor RMS at shock breakout in a stellar wind \citep{Ioka2019, Ito2020}. In this case, the escaping radiation at low optical depths leads to a higher compression behind the shock, which in turn enhances photon production via bremsstrahlung emission. Thus, more photons share the dissipated energy, which decreases the temperature.
However, in the very photon rich regime considered here ($\zeta \sim 10^5$), photon production via bremsstrahlung would remain negligible even at high radiative losses \citep{Ioka2019}. Thus, we obtained the opposite trend for the maximum energy. 

From the figure it is clear that the forward shock upstream had already been depleted at this stage, and the forward RMS had begun to dissolve. Note that the upstream was depleted before the hydro shock reached the front edge, as can be seen from the $\tau = 10$ profile in Figure~\ref{fig:hydro_evolution}. This was because the RMS transition region became broad at low optical depth, and the front of the large transition region reached the front edge of the ejecta earlier than the thin hydro shock. The reverse shock upstream was not yet depleted. Its temperature had dropped by a factor of $\sim 3$ since $\tau = 50$ due to adiabatic cooling. The downstream region contained about half of the total photon population and consisted of a broad spectrum. 
The broad spectrum was partly due to the increasing scattering time, leading to a less efficient thermalisation process, and partly due to the broad reverse shock injecting both lower and higher energy photons than before.

At the photosphere ($\tau = 1$), the downstream contained $\approx 95$\% of all photons. It had a smooth, soft curvature below the peak energy at $\epsilon_p \approx 4\times10^{-3}$ and a clear high-energy component. The high-energy part exhibited a slight spectral hardening before it cuts off sharply, which is a characteristic signature of a photon distribution in the ``fast Compton regime'' \citep{Alamaa2024_intrapulse}.

\begin{figure*}
    \centering
    \includegraphics[width = 0.99\columnwidth]{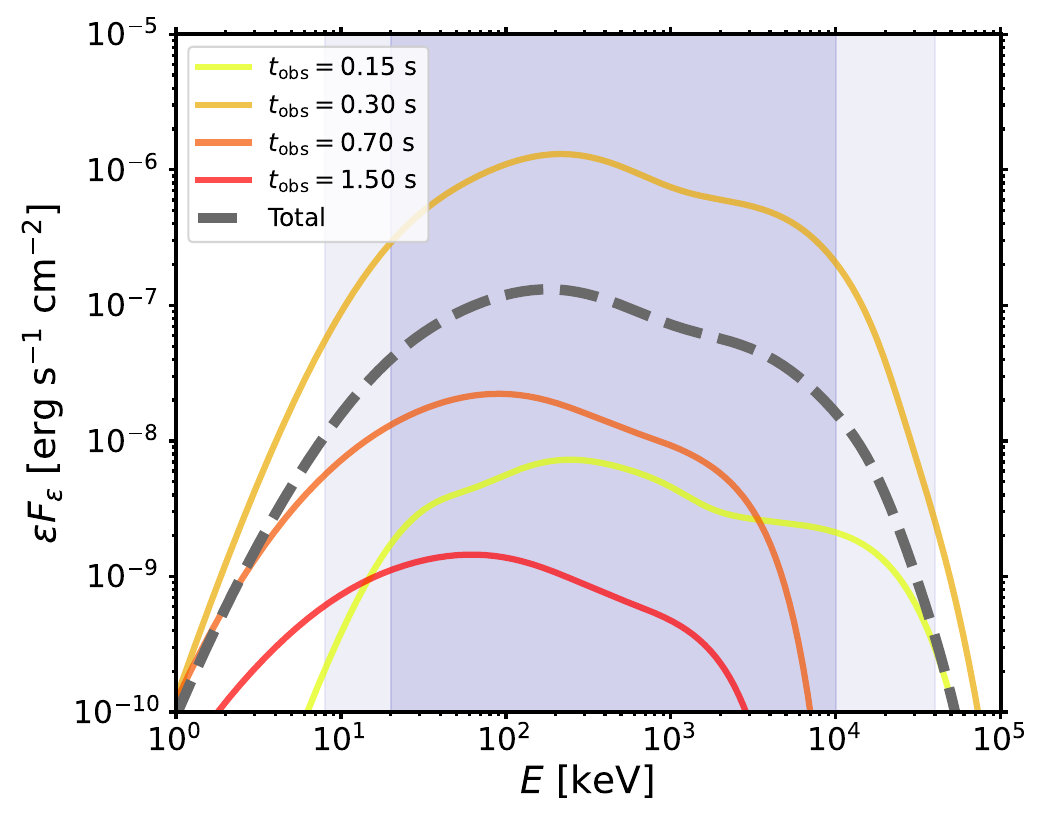}
    \includegraphics[width = 0.99\columnwidth]{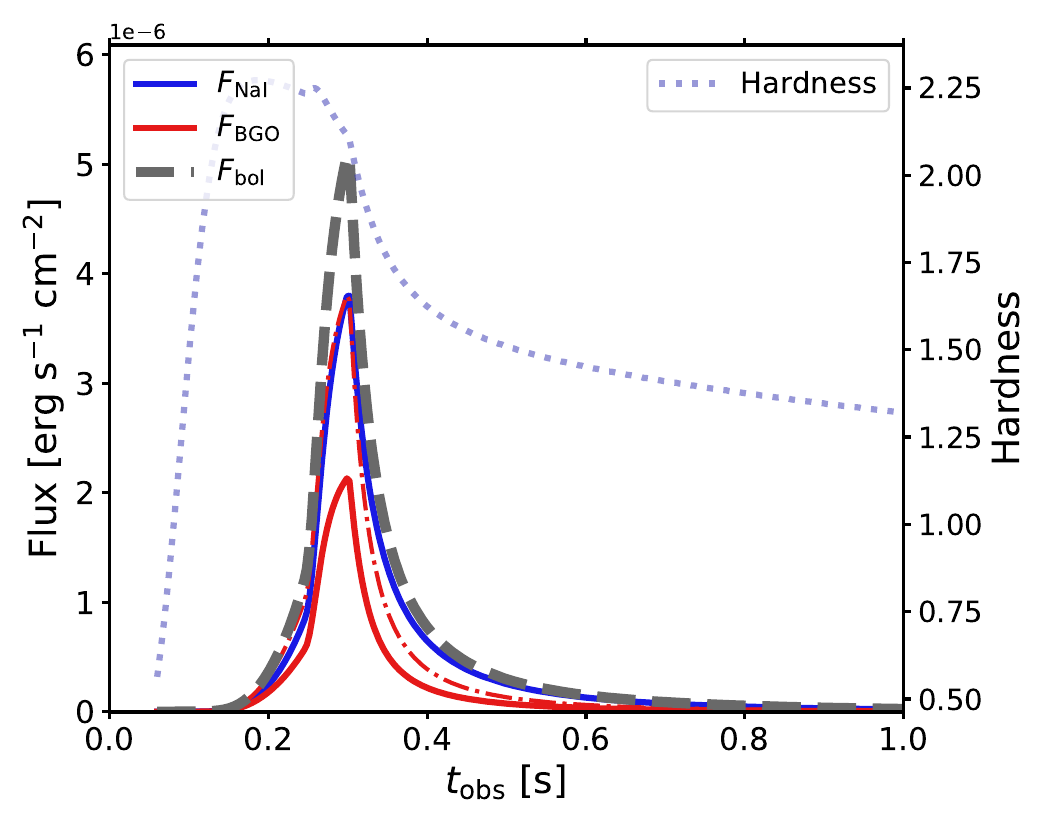}\\
    \includegraphics[width = 0.99\columnwidth]{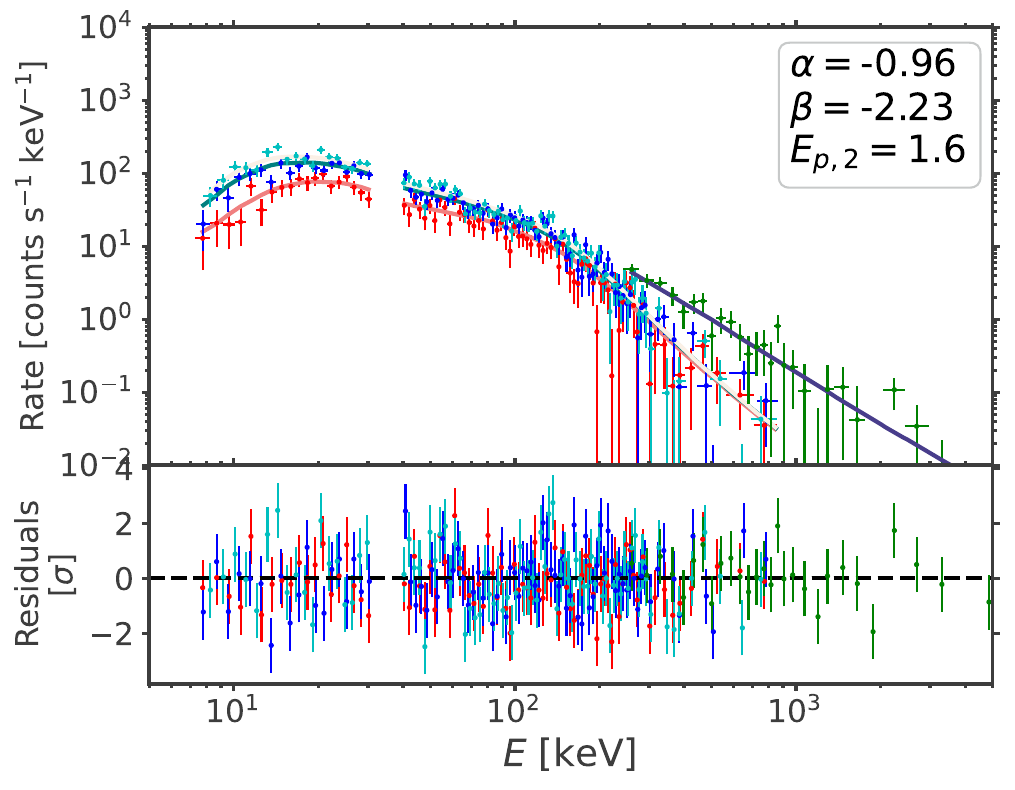}
    \includegraphics[width = 0.99\columnwidth]{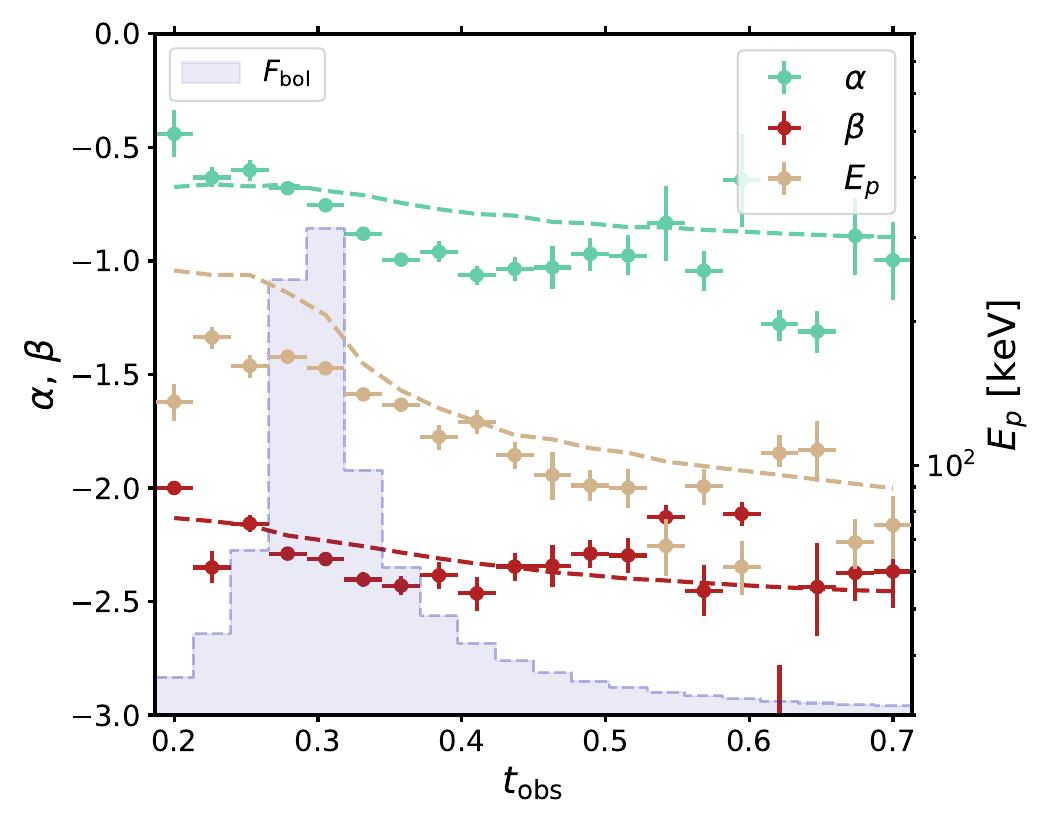}\\
    \caption{Observed spectral evolution (top left), light curve (top right), count spectrum of a Band fit to the time-integrated spectrum (bottom left), and the evolution of the best-fit Band parameters (bottom right). The dashed grey line in the top-left panel shows the time-integrated spectrum taken as the average over the first second of observations, and purple shading shows the energy range of the GBM, with the darker colour indicating the region with the highest effective area. The dot-dashed red line in the light curve shows the light curve in the BGO band scaled to the peak of the NaI light curve to highlight the difference between the two. The red, blue, cyan, and green data points in the count spectrum show the three NaI detectors and the one BGO detector used for the fit, respectively. The best-fit peak energy is given in units $E_{p,2} = E_p/100~{\rm keV}$, and the mock dataset for the fit had a S/N~$=50$. The best-fit parameters with errors for the time-integrated spectrum were $\alpha = -0.96\pm 0.06$, $\beta = -2.23 \pm 0.09$, and $E_p = 156\pm 16~$keV. All errors given represent $1\sigma$. The simulation parameters are given in Table \ref{tab:params}.} 
    \label{fig:standard_case}
\end{figure*}
\begin{figure}
    \centering
    \includegraphics[width = 0.99\columnwidth]{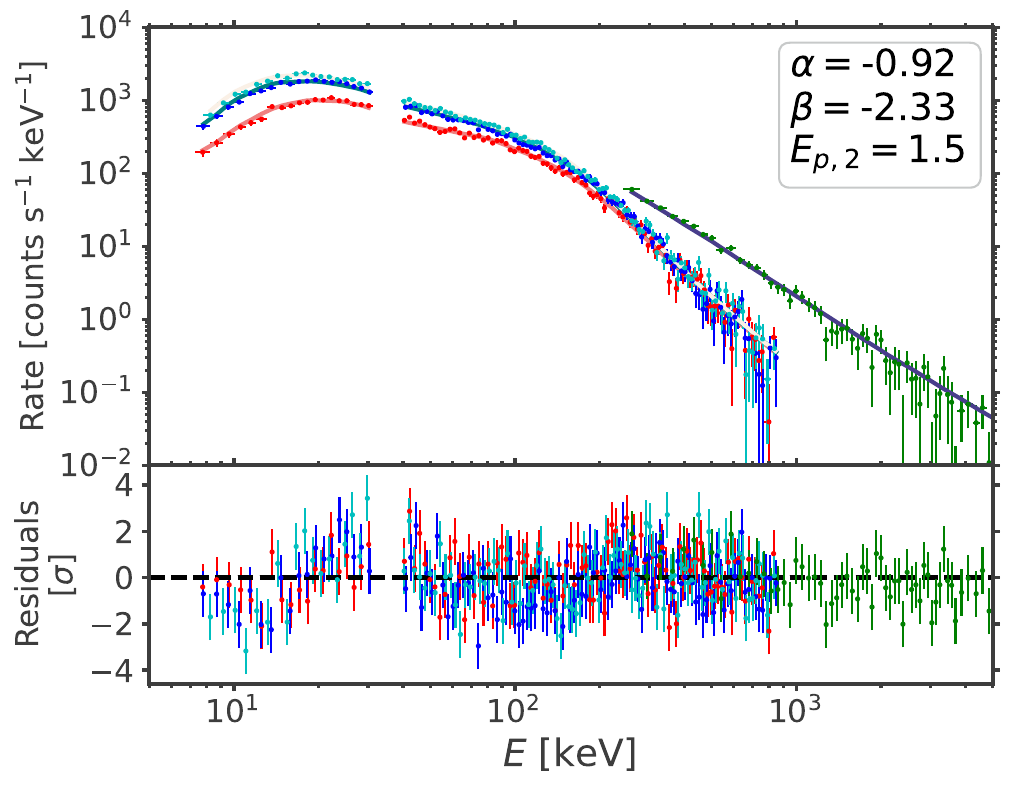}
    \includegraphics[width = 0.99\columnwidth]{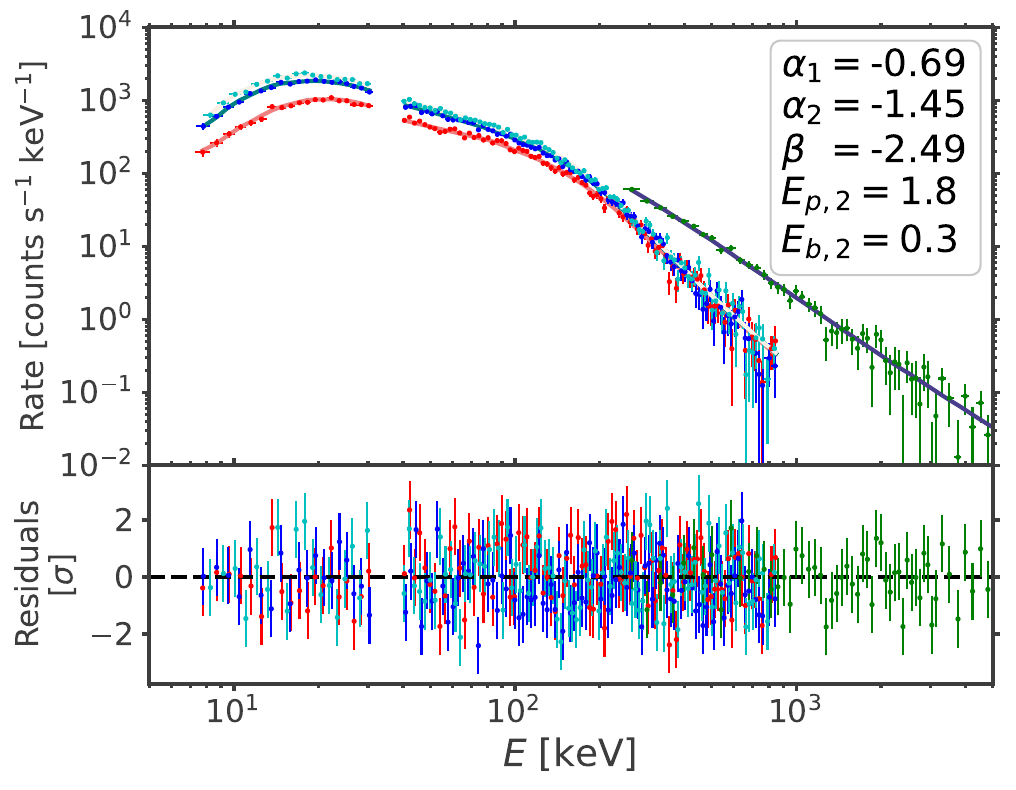}
    \caption{Comparison of a Band function fit (top) and a 2SBPL fit (bottom) to the time-integrated spectrum with S/N~$=300$. The red, blue, cyan, and green data points in the count spectrum show the three NaI detectors and the one BGO detector used for the fit, respectively. It is clear from the uneven residuals in the top panel that the Band function did not produce a satisfactory fit. The 2SBPL function performed much better in this region of high S/N, being statistically preferred with $\Delta {\rm AIC} = 114$. The best-fit Band parameters were $\alpha = -0.92\pm 0.02$, $\beta = -2.33 \pm 0.02$, and $E_p = 152\pm 4~$keV, and the best-fit 2SBPL were $\alpha_1 = -0.69\pm 0.06$, $\alpha_2 = -1.45\pm 0.04$, $\beta = -2.49 \pm 0.04$, $E_p = 178\pm 7~$keV, $E_b = 27\pm 3~$keV.}
    \label{fig:high_SNR}
\end{figure}

\subsection{Light curve}
The observed light curve for a redshift of $z = 1$ is shown in the upper-right panel in Figure~\ref{fig:standard_case}. The grey dashed line shows the bolometric flux. The blue and red solid lines shows the light-curves in the Fermi NaI ($8-1000$~keV) and Fermi BGO ($0.4-40$~MeV) band, respectively, the Fermi NaI and BGO detectors being part of Fermi Gamma-ray Burst Monitor \citep[GBM,][]{Meegan2009}. The energy output in the NaI and BGO bands were comparable and the bolometric flux was completely dominated by the emission within these bands (note that $F_{\rm bol} < F_{\rm NaI} + F_{\rm BGO}$ due to the overlap between the detectors). At peak flux, the burst reached $\sim 5 \times 10^{-6}~$erg~s$^{-1}$~cm$^{-2}$. This is about an order of magnitude higher than the median peak flux of Fermi GRBs \citep[compared to the BEST sample in][]{Poolakkil2021}. Thus, these parameters produced a fairly bright GRB at the current redshift of $z = 1$.

Integrating the bolometric flux, we obtained the radiated gamma-ray efficiency as
\begin{equation}
    \eta_\gamma = \frac{E_{\rm rad}}{E_{\rm tot}} = \frac{\int 4\pi d_l^2 F_{\rm bol} \, \frac{dt_{\rm obs}}{(1+z)}}{\int_0^{t_e} {\dot E} \, dt} \approx 23\%,
\end{equation}
\noindent where the numerator has been integrated over the first second of observation. The value found is quite high. RMSs are very efficient at converting kinetic energy into radiation energy since almost all dissipated energy is taken by the photons. This is different compared to collisionless shocks where the radiation ends up with no more than the fraction of dissipated energy given to electrons, which is usually of the order $\sim 0.1$ \citep{SironiSpitkovsky2011}. However, in contrast to collisionless shocks, the radiation downstream of an RMS suffers from adiabatic degrading, which can severely decrease the radiated gamma-ray efficiency. That the reverse shock terminates at low optical depth was the reason for the high $\eta_\gamma$ in this case. As a consistency check, we found that kinetic energy at the photosphere $E_{\rm kin} (R_{\rm ph}) = \Gamma(R_{\rm ph}) M_{\rm tot}c^2$ was $\approx 79\%$ of the total energy, where $M_{\rm tot} = {\dot M}t_e$ is the total ejected mass, implying that the kinetic plus radiated energy was equal to $E_{\rm tot}$ within $2\%$. This means that we did not artificially inject too much energy into the photons during the RMS modelling. 

The duration of the pulse was short, with a duration comparable to the ejecta duration, $t_e$. It was of the same order as the observed geometrical timescale at the photosphere $t_{\rm geo} = (1+z) R_{\rm ph}/2c\Gamma^2$, which, for the obtained $R_{\rm ph} \approx 3\times10^{13}~$cm and $\Gamma(R_{\rm ph}) \approx 130$, gives $t_{\rm geo} \approx 0.06~$s. 
In this paper, we focused on a single internal collision. A more complex environment would lead to a more complex light curve consisting of many, potentially overlapping pulses, (see further discussion in Section \ref{subsec:disc_complicated_light_curve}).
Furthermore, additional surrounding material can keep the downstream region compressed, which increases the photospheric radius and the observed geometrical timescale. While the aforementioned complexities can moderately increase the pulse duration, many GRBs have pulse durations that are much longer than what we found here \citep{GolkhouButler2014}. These pulses require emission at larger radii, potentially originating from shocks above the photosphere.

The light purple dotted line shows the hardness as a function of time, calculated as the ratio between the flux in the $100$--$300~$keV band to the flux in the $50$--$100~$keV band. It rose by a factor of four from very early times to the light curve peak, after which it decayed smoothly. Although the rise in the hardness ratio is drastic, it would likely be difficult to observe as the GRB was very faint at that time. The decay happened once photons from radii above the reverse shock crossing radius started to reach the observer. When the injection of high-energy photons from the reverse shock ceased and adiabatic cooling started to dominate in the comoving frame, the hardness ratio started to decrease (see further discussion in Section \ref{subsec:result_time_resolved}).

The dot-dashed red line shows the BGO flux scaled to peak at the NaI maximum flux to highlight the differences between the two light curves. The light curve was very similar in the two bands up until the peak but decreased more rapidly in the BGO band. The explanation for the faster decay in the BGO band is the same as that for the decreasing hardness ratio.

\subsection{Time-integrated spectrum}
The observed time-integrated spectrum, taken as the average over the first second of observation, is shown by the dashed line in the top left panel of Figure~\ref{fig:standard_case}. Qualitatively, it was quite similar in shape to the total comoving spectrum at the photosphere as shown in the bottom right panel of Figure~\ref{fig:KRA_comoving_photon_distribution}. However, the superposition of radiation emitted at a range of optical depths (different spectral shapes) and angles (different Doppler boosts) led to a time-integrated spectrum that was broader and softer at low energies and had a more extended high-energy power law. The spectrum peaked at $E_p \approx 200~$keV, which is a typical value for GRBs \citep{Poolakkil2021}. At low energies, the spectrum had a smooth curvature from $E_p$ down to $\sim 3~$keV, below which the spectrum became very steep \citep[tending asymptotically towards $\alpha \sim 0.4$,][]{Beloborodov2010,Lundman2013}. Above the peak, the spectrum exhibited a high-energy power law extending for $\sim 1.5$ orders of magnitude with a slope $\beta \approx -2.3$ cutting off at $\sim 5~$MeV.

To get a more qualitative idea of how such a spectrum compared to actual observations, we used a similar analysis to \citet{SamuelssonRyde2023, Alamaa2024_intrapulse}. The time-integrated spectrum was forward-folded through a detector response matrix. 
This generated a mock dataset that accounted for both detector and background effects. The background was obtained by using the response file from a real GRB, in this case GRB 150213A, and the background strength was determined by setting a user specified signal-to-noise ratio (S/N). The mock dataset could then be fitted with any standard GRB function such as the Band function \citep{Band1993}. In this paper, we used the Fermi Gamma-ray Burst Monitor \citep[GBM,][]{Meegan2009} as the detector. The procedure automatically accounted for the detector effective area, which meant that the fit could only consider data within the GBM energy sensitivity window \citep[8~keV to 40~MeV,][]{Meegan2009}. 

A Band function fit to the time-integrated spectrum is shown in the bottom-left panel in Figure~\ref{fig:standard_case}. The top shows the count spectrum and the bottom shows the residuals. The mock dataset had S/N~$= 50$ and the best-fit parameters for the fit were $\alpha = -0.96\pm 0.06$, $\beta = -2.23 \pm 0.09$, and $E_p = 156\pm 16~$keV where $\alpha$, $\beta$, and $E_p$ are the low-energy spectral index, the high-energy spectral index and the peak energy, respectively. 

It is clear from the residuals in the bottom panel that the Band function was a good fit when S/N~$= 50$. However, the time-integrated spectrum had more complexity below the peak than a single power law. To test this, we generated a new mock dataset with S/N~$= 300$. The Band function fit to this dataset is shown in the top panel of Figure~\ref{fig:high_SNR}. The evident oscillations in the residuals in this case indicated that the Band function was not a good fit at this high S/N. In contrast, a phenomenological function that allows for two breaks below the peak \citep[double smoothly broken power law (2SBPL),][]{Ravasio2018} fitted the spectrum very well. Using the Akaike Information Criterion \citep[AIC,][]{Akaike1974} to evaluate the performance of each model, we found $\Delta {\rm AIC} = {\rm AIC}_{\rm Band} - {\rm AIC}_{\rm 2SBPL} = 114$, indicating that the 2SBPL was greatly favoured by the data. In contrast, we found no statistically significant improvement using the more complex 2SBPL function when S/N~$=50$.

It is interesting to note that the values for the two low-energy slopes were very close to the theoretical values expected in synchrotron emission models, $\alpha_1 = -0.69\pm 0.06$, $\alpha_2 = -1.45\pm 0.04$. As pointed out already in \citet{SamuelssonRyde2023}, there is no a priori reason for why this would be the case. Rather it is a consequence of fitting a function that allows for two power laws below the peak to a curved spectrum that, when fitted with the Band function, produces $\alpha \sim -0.9$. Then, for most fits, one gets $\alpha_2 < \alpha < \alpha_1$, which may by chance give $\alpha_2 \approx -3/2$ and $\alpha_1 \approx -2/3$. This shows that one must be cautious in the interpretation of the underlying physics based on empirical fits.

\subsection{Time-resolved spectrum and spectral evolution}\label{subsec:result_time_resolved}
The instantaneous spectral flux at four different observer times are shown in the top-left panel in Figure~\ref{fig:standard_case}. It is evident that the average energy of the observed radiation decreased with time. Photons that make their last scattering at small radii outrun the ejecta, travelling with a slightly larger velocity compared to the bulk plasma. This means that photons from deeper layers and on the line-of-sight are the first to arrive at the observer. These photons have a high Doppler boost and have suffered less from adiabatic cooling, and therefore they have a higher average energy compared to the radiation that arrives at later times.

The very early observation ($t_{\rm obs} = 0.15~$s, yellow line) had a more complex spectral shape compared to the others, consisting of a main component peaking at $\sim 300~$keV with several additional bumps. 
The radiation observed at early times during the light curve pulse rise come from the same local region in the jet \citep[$\tau \sim 10$, $\mu^\prime \sim 1$,][]{Alamaa2024_intrapulse}. Thus, the observed spectrum at this time should be very similar to the Doppler boosted comoving spectrum at $\tau = 10$. This can be confirmed by comparing it to the total comoving photon distribution at $\tau = 10$ shown in the bottom-left panel in Figure~\ref{fig:KRA_comoving_photon_distribution}. The comparison shows that the main component was due to shocked photons sitting in the downstream region, while the highest energy component originated from the reverse shock. 

The peak flux spectrum ($t_{\rm obs} = 0.3~$s, orange line) was quite similar to the time-integrated spectrum, with a slightly higher peak-energy, a harder low-energy slope, and a curvier high-energy part with a bump-like signature at 5~MeV. At later times, the peak energy decreased and the low-energy part became softer. The high-energy cut-off decreased due to adiabatic cooling, but also due to the depletion of high-energy photons in the comoving frame. Once the injection of fresh photons from the reverse shock ceased, the high-energy part quickly downscattered on the colder electrons, leading to a decrease of the cut-off energy in the comoving frame.

To get a more clear view of the parameter evolution across the GRB, we forward-folded twenty instantaneous spectra through the Fermi GBM response matrix and fitted the mock datasets with the Band function. The resulting best-fit parameters and their uncertainties are shown in the bottom-right panel of Figure~\ref{fig:standard_case}. The S/N was set to be 300 for the peak flux bin, while the other bins had their S/N determined by their relative brightness. The high maximum S/N was chosen such that the fitted parameter evolution could be seen more clearly. However, we note that a time-resolved analysis with twenty bins of such high quality in a short pulse similar to the one studied here would be extremely rare.

The fitted values are compared to theoretical estimates showed by dashed lines. The theoretical peak energy, $E_{p, {\rm th}}$, was taken to be the maximum of the $\epsilon F_\epsilon$ spectrum. The theoretical indices were estimated as $\alpha, \beta = d\log F_\epsilon/d\log\epsilon - 1$ evaluated at $0.1E_{p, {\rm th}}$ and $10E_{p, {\rm th}}$ for $\alpha$ and $\beta$, respectively.

The fitted low-energy index decreased from $\alpha \approx -0.5$ to $\alpha \approx -1.0$ between $0.2~$s and $0.4~$s. After $0.4~$s, it remained $\sim~$constant although all fitted values showed a large scatter at $t_{\rm obs} > 0.5~$s due to the low S/N. Comparing with the theoretical estimate, we can see that the fit generally underestimated the slope at $0.1E_p$. This was because the slope just below the peak was softer and since it was the brightest part of the spectrum, it dominated the fit. At late times, the hardest low-energy curvature was below the GBM detector window, which also led to the fit underestimating the value of the slope. This tendency becomes more and more pronounced as the peak energy decreases further \citep{Acuner2019}.

The fitted values of the peak energy increased slightly during the pulse rise time. However, it did not represent the real peak energy very well. 
This was due to the complex spectral shape that did not resemble a Band function at early times. The real peak energy had a high ($\sim 300~$keV), constant value during the pulse rise, after which it started decreasing due to adiabatic cooling. That the peak energy was not decreasing during the pulse rise time was because of the existence of the reverse shock at low optical depths. Unless there is some ongoing dissipation in the decoupling region ($\tau < 10$), the peak energy decreases throughout the pulse \citep{Alamaa2024_intrapulse}.

The fitted values of the high-energy index decreased during the pulse rise time from $\beta= -2$ to $\beta = -2.4$ and remained $\sim$constant thereafter, albeit with a large scatter. Although subtle, such a signature is quite revealing and arises from photons with high comoving energies being downscattered in the comoving frame. Since the Compton cooling is proportional to the photon energy, these high-energy photons are the only ones that have time to be downscattered in the decoupling region. Energy transfer via Compton scattering is only effective during the pulse rise time before adiabatic cooling becomes the dominant energy loss channel \citep{Alamaa2024_intrapulse}. One of the ways that this can manifest itself in the data is by a rapid decrease of the high-energy index during the pulse rise followed by a constant value during the pulse decay.

\section{Discussion}\label{sec:discussion}

\subsection{Single pulse vs highly variable light curve}\label{subsec:disc_complicated_light_curve}

In this paper, we focused on a single internal collision with an idealised initial Lorentz factor profile. This simple setup is unlikely to be realistic for a real GRB. However, investigating a single collision is a good starting point before moving on to more realistic, highly variable ejecta. 

The single collision considered here produced a light curve consisting of a $\sim 0.1~$s duration pulse. This pulse could be one building block within a complicated, highly variable longer GRB light curve. Numerical simulations of collapsar GRBs show that multiple shocks usually form within the relativistic outflow \citep[``shock train'', e.g.][]{Lopez-Camara2014, Ito2015, Ito2019, Gottlieb2019, Gottlieb2020}. Specifically, \citet{Gottlieb2019, Gottlieb2020} found that mixing around the collimation shock can result in strong short term variability for the terminal Lorentz factor, with typical values close to what we considered here. This variability led to internal collisions, some of which occured below the photosphere \citep{Gottlieb2019}.

If the single collision considered here occurred within a broader variable ejecta with multiple other shocks, the periodic boundary condition used by the second simulation performed are more appropriate (see Section \ref{subsec:boundary_conditions}).
Remarkably, we found that the results from that simulation were very similar to those presented for the vacuum boundary case in Section \ref{sec:results}.\footnote{The additional material outside of the boundary was accounted for when calculating $\tau$ by assuming a periodic $\rho R^2$.} The boundary conditions only became important at low optical depth when the shocks had crossed the considered region of the ejecta, and therefore, they had little effect on the radiation. Furthermore, the optical depth itself was not much affected either, since the previous high-velocity material had left a low-density cavity in its wake, meaning that photons made their last scattering at roughly the same radii as in the vacuum case. We conclude that the results presented in Section \ref{sec:results} can be applied to short pulses in complex light curves as well, while long duration pulses may require shocks occurring above the photosphere. Note that the spectral evolution obtained here would then be for a single pulse within the light curve; the complex GRB as a whole can have a vastly different evolution \citep[see e.g.][]{Ito2019}.

The light curve shown in Figure \ref{fig:standard_case} could also constitute a whole short GRB, as expected in a merger event. For instance, \citet{Pais2024} simulated short GRBs by injecting high power jets into a realistic neutron star merger environment \citep[obtained from a 1~s long 3D GR neutrino-radiation MHD simulation,][]{Kiuchi2023}. In one of their simulations, they found that the jet material clustered in velocity space around $\Gamma \approx 40$ and $\Gamma \approx 400$ at the time of jet break out. The slower material consisted in part of jet material shocked by the jet head reverse shock as it propagated through the neutron star debris, while the faster material, situated further inwards, consisted of unshocked material. This would lead to the formation of shocks at a larger distance and gives some support to the Lorentz factor profile used in this paper.

\subsection{Non-thermal photospheric emission from RMSs}\label{subsec:disc_high_efficiency}
Dissipative photospheric models require that the dissipation occurs not too deep below the photosphere for the emitted spectrum to be non-thermal \citep[outside of the Wien zone, $\tau \lesssim 100,$][]{Beloborodov2013}.
In the current framework, there are two physical reasons for why RMSs that start at high optical depth will have a tendency to finish their dissipation close to the photosphere. The first is that while both forward and reverse shocks exist, the downstream stays compressed, keeping the optical depth high. Once either shock has crossed, the high internal energy in the downstream region leads to an acceleration of the ejecta (a second fireball), which rapidly decreases the comoving density and, subsequently, the optical depth. Thus, it is plausible that the transition to transparency happens quickly after the RMS has crossed.

The second is that number of photons in the RMS transition region increases when the shock reaches low optical depths \citep[][see also Appendix~\ref{App:shock_broadening}]{Levinson2012}. Indeed, it is clear from the approximate expression in Equation~\eqref{eq:dN_in_dr} that the photon number increases exponentially when $\tau/{\mathcal N}_{\rm sc} \ll 2$, where ${\mathcal N}_{\rm sc}$ is the average number of scatterings a photon performs in the RMS. This means that the shock front effectively ``swallows'' the upstream when the shock comes close to the photosphere. To what extent this latter phenomenon would occur in reality needs to be checked with more detailed numerical simulations.

\subsection{High-energy power law}
The time-integrated spectrum had a high-energy power law that extended for $\sim 1.5$ orders of magnitude before it cut off at $\sim 5~$MeV. It was generated by the injection of high-energy photons from the shock into a region of relatively cold plasma, where they suffered Compton losses \citep{Alamaa2024_intrapulse}. However, it is not straight forward in photospheric models to get a high-energy power law extending beyond $\sim \Gamma m_e c^2$ due to the high $\gamma\gamma$-opacity \citep[although see][]{Beloborodov2010}. Thus, it is difficult to account for the sample of GRBs that have a power law extending above several hundred MeV as those presented in \citet{Ravasio2024, Macera2025}. Potentially, such high-energy photons could be generated at the transition from an RMS to a collisionless shock around the photosphere.

\subsection{Neglecting radiation feedback}\label{subsec:disc_radiation_feedback}
In this paper, we have used a pure hydrodynamical simulation to get a handle on the GRB ejecta dynamics. However, below the photosphere, the photons dominate the pressure, meaning that the hydrodynamical profile will be determined by the radiation. For instance, the detailed shock structure is dependent on the photon scattering length and it can only be accurately determined if radiation feedback on the hydrodynamics is included. 
However, in this paper, the hydrodynamical simulation was used to obtain general properties regarding the shocks, such as the compression, the downstream enthalpy, and the downstream Lorentz factor. Such global quantities are determined by the shock jump conditions, which must be satisfied regardless of the shock microphysics. The photon distributions in the two RMSs were then modelled by the KRA, which has been compared to detailed numerical simulations where radiation feedback was included with good results in \citet{Samuelsson2022}.

The above described picture changes at low optical depths. Close to the photosphere, photons decouple, which means a loss of internal energy that is currently unaccounted for. Indeed, photon escape changes the dynamics of RMSs as shown in \citet{Ito2020, Ito2026}. Here, both shocks had crossed at moderately thick optical depths ($\tau > 7$), where photon escape is not yet highly significant \citep{Beloborodov2011}. Thus, we argue that the main conclusions are unaffected by this approximation.

\subsection{Klein-Nishina and pair production}\label{subsec:disc_KN_pair}

Klein-Nishina (KN) effects and pair production becomes important once the photon energies in the comoving frame become comparable to the electron rest mass. Both of these effects were neglected here. Including KN effects, the spectrum should soften above $\epsilon \gtrsim 0.1$ \citep{Ito2018, Lundman2018}, meaning that the simulation overestimated the number of photons at the highest energies. This would have in turn affected the detailed shape of the high-energy part of the spectrum. For instance, the high-energy bump visible at $\gtrsim10~$MeV at $t_{\rm obs} = 0.15~$s in Figure~\ref{fig:standard_case} would have become less pronounced or disappeared. Furthermore, there would have been a precursor of high-energy photons in the light curve at energies above $\sim 10 \frac{\Gamma_2}{(1+z)}~$MeV, where $\Gamma_2 = \Gamma/100$. However, KN effects would likely not have had a big impact on the estimated best-fit parameters, which were determined by the spectral shape around the comoving peak energy at $\epsilon_p \approx 4 \times 10^{-3}$ where KN effects are negligible.

RMSs can lead to violent pair production if the shock velocity is high \citep{Budnik2010, Katz2010}.
In the current simulation, very few photons had energies high enough for pair production. In addition, this number was overestimated as KN effects were ignored. Furthermore, the spectrum of the reverse shock at $\tau = 10$ was quite similar to a detailed RMS simulation performed in \citet{Ito2018} (Figure~4, bottom-left panel, similar maximum energy and similar extension above $\epsilon > 1$), which was found to produce no pairs in the RMS transition region. 
We conclude that including KN effects and pair production would not have altered the main conclusions of the paper. 

\section{Conclusion}\label{sec:conclusion}
In this paper, we have used a series of three numerical codes to go from initial parameter distributions deep below the photospheric radius in a GRB jet all the way to the best-fit parameter evolution of the time-resolved observed signal (see Figure~\ref{fig:flowchart} for a flowchart of the methodology). A 1D special relativistic hydrodynamical simulation code \citep[\texttt{GAMMA},][Section~\ref{sec:hydro_implementation}]{Ayache2022} was used to model an internal collision that generated a reverse and a forward shock below the photosphere. The simulation was set up such that both shocks finished their dissipation in the optically thick regime close to the photosphere. The evolution of the Lorentz factor and the comoving density as a function of optical depth was shown in Figure~\ref{fig:hydro_evolution}.

Since the shocks were subphotospheric, they were mediated by radiation, and thus a designated RMS model was needed. For this purpose, we used the Kompaneets RMS Approximation (KRA) developed in \citet{Samuelsson2022}. However, to be able to model long-lasting, dynamically evolving shocks, and to be able to account for the contribution from both the reverse and the forward shock, several additions and modifications had to be made. This led to the development of the dynamical version of the KRA, summarised in Section~\ref{sec:KRA_dynamical} and described in detail in Appendix~\ref{app:KRA_dynamical} (see Figure~\ref{fig:KRA_geometry} for a schematic of the geometry). The dynamical version of the KRA was modelled by the simulation code \Kd\ and the obtained comoving photon distribution as a function of optical depth was shown in Figure~\ref{fig:KRA_comoving_photon_distribution}. 

Finally, the time-resolved signal was obtained using \texttt{Raylease} \citep{Alamaa2024_intrapulse}. It used the ejecta dynamics from \texttt{GAMMA} and the subphotospheric photon distribution from \Kd \ as input to generate the spectrum as a function of time in the observer frame. \texttt{Raylease} accounts for the probabilistic nature of the photosphere, including emission from different optical depths and angles to the line-of-sight.

The results were shown in Figures~\ref{fig:standard_case} and \ref{fig:high_SNR} and can be summarised as follows:

\begin{enumerate}
    \item The light curve consisted of a single pulse with a sharp peak and had a duration that was comparable to the geometrical timescale at the photosphere, which in this case was $\sim 0.1~$s. It could either constitute a whole GRB in itself or be a part of a more complex longer light curve. The radiative efficiency was high at $\sim 23$\% due to the reverse shock existing at low optical depths. The hardness peaked at the light curve peak, and then slowly decayed due to adiabatic cooling in the comoving frame and high-latitude emission. 
    \item The time-integrated spectrum consisted of a smooth curvature below the peak energy a high-energy power law that extended to $\sim 5~$MeV. For a redshift of $z = 1$ and S/N$~=50$, it had best-fit Band parameters of $\alpha = -0.96\pm 0.06$, $\beta = -2.23 \pm 0.09$, and $E_p = 156\pm 16~$keV. It was well fitted with a Band function in the low S/N region, but was much better fitted with a 2SBPL function in the high S/N region. 
    \item A high-energy power law extending for 1.5 orders of magnitude in energy was generated due to the injection of high-energy photons from the reverse shock. The maximum energy in the reverse shock increased at lower optical depths partly because the shock sped up in the decreasing density profile and partly because the decreasing scattering time led to a steeper velocity gradient across the shock (in units of optical depth).
    \item The peak energy remained $\sim~$constant during the pulse rise due to the existence of the reverse shock. After the reverse shock crossing, adiabatic cooling led to the observed peak energy decreasing. The high-energy index showed the opposite behaviour with a rapid decrease during the pulse rise, after which it stayed $\sim~$constant. The decrease was due to a rapid downscattering of high-energy photons, a process that becomes inefficient at low optical depths ($\tau \lesssim 5)$.
    \item The very early spectrum probed the deeper regions of the jet and showed a more complex behaviour, where both reverse and forward shocks were visible.
\end{enumerate}

In conclusion, we find that RMSs show several promising features with regards to the prompt GRB emission. However, we have only studied a single example in this paper. It remains to be tested if the model can reproduce the diversity of observed light curves and pulse durations, and under which conditions some shocks can still propagate above the photosphere with a different contribution to the emission.

\begin{acknowledgements}
    We thank the anonymous referee for their insightful comments. This research has made use of the High Energy Astrophysics Science Archive Research Center (HEASARC) Online Service at the NASA/Goddard Space Flight Center (GSFC). We thank the GBM team for providing the necessary tools and data. F.A. is supported
    by the Swedish Research Council (Vetenskapsr\aa det; 2022-00347). F.D. acknowledges the Centre National d'\'Etudes Spatiales (CNES) for financial support in this research project.
\end{acknowledgements}

\bibliographystyle{mnras}
\bibliography{References}

\newpage

\begin{appendix}

\section{Shock broadening and the dynamical stage of an RMS}\label{App:shock_broadening}

The width of the RMS is set by the number of scatterings the photons have to do to decelerate the incoming flow. As the photon mean free path increases with time due to the decreasing lepton density in the jet, the physical width of the RMS grows. 

The instantaneous optical depth across the shock transition region, $\Delta \tau_r$, (not to be confused with the optical depth towards the observer, $\tau$) is given by an integral over the shock transition region as
\begin{equation}\label{eq:Delta_tau_r}
    \Delta \tau_r = \int n_{e} \sigma_{\rm T} dR^\prime 
    \approx n_{e} \sigma_{\rm T} \Delta R^\prime_r,
\end{equation}
\noindent where $n_{e}$ is the comoving electron number density,\footnote{The comoving lepton density is going to vary across the shock transition due to compression. The value of $n_{e}$ used here should be viewed as an average value over the shock.} $\sigma_{\rm T}$ is the Thomson cross section, $\Delta R^\prime_r$ is the width of the shock as measured in the downstream comoving frame, and we assumed $\Delta R^\prime_r \ll R$ and a non-relativistic RMS. The number of photons in the shock transition region is given by $N_r = 4\pi R^2 n_{\gamma} \Delta R^\prime_r$, where $n_{\gamma}$ is the comoving photon number density. Thus, the photon number in the RMS is given by
\begin{equation}\label{eq:N_r}
    N_r(R) = \frac{4 \pi R^2 \zeta \Delta \tau_r}{\sigma_{\rm T}},
\end{equation}
\noindent where the photon-to-proton ratio, $\zeta$, is equal to the photon-to-electron ratio in the absence of pairs.\footnote{Since the electron and photon densities vary similarly across the shock due to compression, Equation~\eqref{eq:N_r} is independent of the compression.} The photon-to-proton ratio remains $\sim$\,constant across the photon transition in a photon rich shock.
Equation \eqref{eq:N_r} implies that the photon number in the RMS increases as $N_r \propto R^2$, with instantaneous change in the lab frame as
\begin{equation}\label{eq:dN_r_dr}
    \frac{d N_r}{dR} = \frac{8 \pi R \zeta \Delta \tau_r}{\sigma_{\rm T} } = \frac{2}{R}N_r,
\end{equation}
\noindent assuming that $\zeta$ and $\Delta \tau_r$ vary slowly with $R$.

Lets assume that the probability for a photon to leave the shock per scattering time is inversely proportional to the average number of scatterings in the shock, ${\mathcal N}_{\rm sc}$. 
Then the outgoing photon number is given by
\begin{equation}\label{eq:dN_out_dr}
    \frac{d N_{\rm out}}{dR} = \frac{N_r}{{\mathcal N}_{\rm sc}} (\beta_r ct_{\rm sc})^{-1} \approx \frac{N_r}{{\mathcal N}_{\rm sc}}\frac{\tau}{R},
\end{equation}
\noindent where $\beta_r \approx 1$ is the dimensionless shock speed in the lab frame, and we used $t_{\rm sc} = \Gamma/\rho \kappa c$ and $\tau \approx \rho \kappa R/\Gamma$ for the last approximation. Combining Equations~\eqref{eq:dN_r_dr} and \eqref{eq:dN_out_dr}, we get the ingoing photon number as 
\begin{equation}\label{eq:dN_in_dr}
    \frac{d N_{\rm in}}{dR} = \frac{d N_r}{dR} + \frac{d N_{\rm out}}{dR} \approx \frac{N_r}{R}\left(2 + \frac{\tau}{{\mathcal N}_{\rm sc}}\right).
\end{equation}

From the equation above, it is clear that in the optically thick regime $\tau > 1$, there exists two different limiting cases in the shock evolution. When ${\mathcal N}_{\rm sc}\ll\tau/2$, the outgoing photon number roughly equals the incoming photon number. In a steady outflow with $\rho \propto R^{-2}$, the optical depth is halved during one dynamical time. Thus, when ${\mathcal N}_{\rm sc}\ll\tau/2$, photons cross the RMS well within a dynamical time and the shock has time to reach a steady state in response to the slowly varying upstream properties. However, once ${\mathcal N}_{\rm sc} \sim \tau/2$, the dynamical stage of the shock evolution starts. At this stage, the shock crossing timescale equals the dynamical time and the shock will struggle to respond to the changing upstream properties. Due to adiabatic cooling in the upstream, the average number of scatterings in the shock is not enough to dissipate the necessary energy and decelerate the incoming flow. The velocity gradient (in units of photon mean free paths) across the shock must steepen, which in turn will increase the average energy gain per scattering for the photons until the shock jump conditions are satisfied. If the shock jump conditions are not satisfied even with the steeper velocity gradient, a collisionless subshock must form \citep{Ito2020, LundmanBeloborodov2021}. Note that at this stage, $\Delta \tau_r$ is no longer slowly varying with $R$ and Equation \eqref{eq:dN_in_dr} breaks down.

For non-relativistic RMSs, the number of scatterings follows ${\mathcal N}_{\rm sc} \propto \beta_u^{-2}$ \citep{BlandfordPayneII1981}. This implies that the transition to a dynamically evolving shock can happen deep below the photospheric radius, much earlier than the shock breakout time. 

The same result can be obtained using Equation~\eqref{eq:Delta_tau_r}, the approximate scaling $\Delta \tau = \beta_u^{-1}$ \citep[e.g.][]{LevinsonNakar2020}, and $n_e = \Gamma \tau/\sigma_{\rm T}R$ valid for a steady wind in the absence of pairs. Then, one finds that
\begin{equation}
    t_{\rm cross} = \frac{t_{\rm dyn}}{\beta_u^2 \tau},
\end{equation}
where $t_{\rm cross} = \Delta R^\prime_r/\beta_u c$ is the shock crossing time and $t_{\rm dyn} = R/\Gamma c$ is the dynamical time. Thus, the condition $t_{\rm cross} \ll t_{\rm dyn}$ requires $\tau \gg 1/\beta_u^2$. Shock breakout occurs once the optical depth towards the observer is of the order $\Delta \tau_r$, which means that at breakout, the crossing time is a factor $\beta_u^{-1}$ larger than the dynamical time according to the equation above.

\section{Obtaining the RMS parameters from the hydro output}\label{App:hydro_to_RMS}
\subsection{Upstream, immediate downstream, and far downstream properties}\label{subsec:H_obtaining_properties}
The hydrodynamical simulation produced two shocks, which separated the ejecta into five regions (see Section~\ref{subsec:H_geometry} and Figure~\ref{fig:KRA_geometry}).
\texttt{GAMMA} comes with its own shock finding 
algorithm. However, here we instead used Equation (72) in \citet{MartiMuller1996} \citep[see also][]{ColellaWoodward1984}, which considers cell $j$ to be within a shock if the relative pressure increase across the cell satisfies
\begin{equation}\label{eq:shock_detection}
    \frac{|p_{j+1} - p_{j-1}|}{\min(p_{j+1}, p_{j-1})} > \delta,
\end{equation}
where $\delta$ is a constant that we set to 1 \citep[in accordance with ][]{MartiMuller1996}. Equation~(72) in \citet{MartiMuller1996} also includes a velocity condition that has been neglected here, since we wanted to identify both left-moving and right-moving shocks.

Since the boundary condition could lead to spurious shock detections close to the boundary, Equation~\eqref{eq:shock_detection} was only evaluated between Lagrangian mass coordinates $m_{\rm ll} = 0.01M_{\rm tot}$ and $m_{\rm ul} = 0.99M_{\rm tot}$.

Numerically, the discontinuity corresponding to a shock is always smoothed over a few cells. Therefore, Equation~\eqref{eq:shock_detection} identified several consecutive cells as being within the shock, for both the forward and the reverse shock (of the order of three-four cells for the used grid resolution). The upstream properties in each saved snapshot were taken to be the properties in the cell that was five cells upstream of the shocked cells. Similarly, the immediate downstream properties were taken at five cells downstream of the shocked cells. As an example, say that Equation~\eqref{eq:shock_detection} identified cells 293, 294, and 295 as within the reverse shock and cells 741, 742, 743 and 744 as within the forward shock. The upstream (immediate downstream) properties for the reverse shock and the forward shock were then taken to be the properties of cell 288 (300) and cell 749 (736), respectively. The choice of five was found to give a good representation of the pressure, density, and velocity in the upstream and the immediate downstream for the current grid resolution.

For the properties in the downstream region, we set the value of a quantity $Q$ at time $t$ to equal the mass-averaged value between the two shocks as
\begin{equation}
    Q(t) = \frac{1}{M_{\rm tot}}\int Q(t,m) \, dm,
\end{equation}
\noindent where the lower (upper) limit to the integral is the mass coordinate of the reverse (forward) shock at time $t$. The bulk Lorentz factor, the scattering time, and the adiabatic cooling term for the downstream region were all calculated this way.

\subsection{Obtaining the RMS parameters}\label{subsec:H_KRA_RMS_parameter_evolution}

Using the shock jump conditions from \citet{Beloborodov2017} \citep[see also][]{Samuelsson2022}, we solved for the upstream Lorentz factor of the incoming material in the shock rest frame as
\begin{equation}\label{eq:gamma_u}
    \gamma_u^2 = \frac{1 - (\rho_u / \rho_d)^2}{(1+w_u)^2/(1+w_d)^2 - (\rho_u / \rho_d)^2},
\end{equation}
where $w = 4p/\rho c^2$ is the dimensionless enthalpy obtained directly from the hydro.

The photon-to-proton ratio was estimated by assuming a thermodynamic equilibrium at $R_0$, a negligible energy in magnetic fields compared to the internal energy of radiation and pairs at $R_0$, and that all pairs had recombined at the collision radius. This gives \citep{Bromberg2011b, LevinsonNakar2020, Alamaa2024_intrapulse} 
\begin{equation}\label{eq:zeta}
    \zeta = \frac{1}{4\theta_0} \frac{\eta}{\Gamma_0} \frac{m_p}{m_e},
\end{equation}
where $\theta_0$ is the temperature at $R_0$ in units electron rest mass given by
\begin{equation}
    \theta_0 = \frac{k_{\rm B}}{m_e c^2} \left[\frac{3}{11} \frac{{\dot E}}{4 \pi R_0^2 a c \Gamma_0^2 \beta_0}\right]^{1/4}.
\end{equation}
In the equation above, $a$ is the radiation constant and $\beta_0 = \sqrt{1 - \Gamma_0^{-2}}$. Since $\zeta$ depends on $\eta$ and ${\dot E}$, it varied across the ejecta. However, as given in Equation~\eqref{eq:zeta}, it remains constant in time for a given mass coordinate. Therefore, in our Lagrangian framework, each cell had one constant value of $\zeta$ associated with it. This is an approximation, as photons would flow between different mass elements in reality. However, in the high optical depth regions treated here, photons are tied to the plasma and this approximation was likely quite good.

Since the photons completely dominated the particle number, the upstream temperature in units electron rest mass energy was given by $\theta_u = p_u/n_{u, \gamma} m_e c^2$. Thus,
\begin{equation}\label{eq:theta_u}
    \theta_u = \frac{w_u}{4}\frac{m_p}{m_e} \frac{1}{\zeta_u}.
\end{equation}
\noindent Finally, the average photon energy in the immediate downstream in units electron rest mass energy was obtained from the relation ${\bar \epsilon}_d = 3p_d/n_{d,\gamma} m_ec^2$ where we used a relativistic equation of state and assumed that the photons dominate the downstream pressure. This gives
\begin{equation}\label{eq:eps_d}
    {\bar \epsilon}_d = \frac{3w_d}{4}\frac{m_p}{m_e} \frac{1}{\zeta_d}.
\end{equation}
\noindent The downstream average energy, ${\bar \epsilon}_d$, can be obtained from $\gamma_u$, $\zeta_u$, and $\theta_u$ using the shock jump conditions, i.e. it is not a free parameters if the other three are known.

\section{Details regarding the KRA in a dynamical flow}\label{app:KRA_dynamical}
In this appendix, the interested reader can find all necessary details regarding the dynamical version of the KRA.

\subsection{Geometry of the KRA with a reverse and a forward shock}\label{subsec:KRA_geometry}

The original version of the KRA only treated one RMS. However, the dynamical version of the KRA can model both the forward and the reverse shock resulting from a collision. It is possible to run the dynamical version of the KRA with only one RMS as well. In other words, all the equations written in this appendix are applicable in the case of either one or two RMSs, unless otherwise stated. Furthermore, the reverse and forward shocks are modelled identically. Thus, we do not generally state which shock a specific quantity, such as $\theta_{u, {\rm K}}$ or $\theta_r$, belongs to.

The material flowing through the forward and reverse shocks is situated in a common downstream with $\sim~$constant pressure and velocity. The observed width of the common downstream marginally satisfies $\Delta R \lesssim R/\Gamma^2$ \citep{Levinson2012}, which means that the downstream light crossing time is shorter than the dynamical time. Therefore, we choose to treat the photons in the downstream as a single fluid, regardless of which shock they originally crossed. 
For the simulation presented in Section~\ref{sec:results}, the condition $\Delta R \lesssim R/\Gamma^2$ broke down when $\tau < 2$. However, at these small optical depths, the photon distribution changes mostly due to adiabatic cooling and the single zone approximation makes no difference.

To evolve both the forward and the reverse shock, the KRA has five different zones: 
1) the reverse shock upstream zone, 2) the reverse shock zone, 3) the common downstream zone, 4) the forward shock zone, and 5) the forward shock upstream zone. 
This geometry is shown in Figure~\ref{fig:KRA_geometry}.

In each zone, photons interact with the plasma electrons via thermal Comptonisation. The photons in the upstream zones are assumed to be in a thermal Wien distribution at the corresponding upstream temperature, $N_{u,\epsilon} \propto \epsilon^2 \exp(-\epsilon/\theta_{u, {\rm K}})$. The effect of the dissipation of kinetic energy in a true RMS is modelled by enforcing the electrons in the shock zones to have a high effective temperature.
The downstream temperature is assumed to equal the Compton temperature, which is analytically obtained from the downstream photon distribution, $N_{\ast,\epsilon}$, as \citep{BeloborodovIllarionov1995}

\begin{equation}\label{eq:theta_C}
    \theta_{\ast, {\rm C}} = \frac{1}{4}\frac{\int \epsilon^2 N_{\ast,\epsilon} \, d\epsilon}{\int \epsilon N_{\ast, \epsilon} \, d\epsilon}.
\end{equation}
\noindent Since $\theta_{\ast, {\rm C}}$ is a function of the downstream photon distribution, it needs to be continuously updated as the distribution evolves.

The different zones are coupled via source terms as indicated in Figure~\ref{fig:KRA_geometry}. These are given in subsection~\ref{subsub:source_terms}.

\subsection{New version of the Kompaneets equation}\label{subsec:KRA_sph_varying_Gamma2}

In \citet{Samuelsson2022}, the Kompaneets equation was written as a function of the normalised radius ${\bar R} = R/R_{\rm ph}$. This is not a suitable choice when the downstream properties, such as the bulk Lorentz factor, can vary, since the estimated value of the photospheric radius will change over time. Thus, we rewrite the Kompaneets equation as a function of $R$. Using the generic subscript $z$ to refer to an arbitrary KRA zone, and neglecting induced scatterings and photon creation and absorption, it can be written as \citep{VurmBeloborodov2016}

\begin{equation}\label{eq:Kompaneets_sph}
\begin{split}
	& \frac{\partial {\mathcal F}_z}{\partial R} = \frac{s_z(\epsilon, R)}{\epsilon^2} + \\
    &\frac{1}{\epsilon^2}\frac{\partial}{\partial \epsilon}\left[ \frac{\epsilon^4 }{t_{{\rm sc}, z}c}\left( \theta_z \frac{\partial {\mathcal F_z} }{\partial \epsilon} + {\mathcal F_z} \right) + \varphi_z \frac{\epsilon^3 {\mathcal F_z}}{R} \right].
\end{split}
\end{equation}
\noindent In the equation above, $s_z$ is a source term (further discussed in the next subsection) and ${\mathcal F}_z$ is related to the photon spectral number density in zone $z$ as ${\mathcal F}_z \equiv N_{z, \epsilon}/\epsilon^2$.
The total number of photons in the zone is given by
\begin{equation}\label{eq:N_z}
    N_z = \int_0^\infty N_{z,\epsilon} \, d\epsilon.
\end{equation}
The Kompaneets equation is solved using the numerical scheme outlined in \citet{ChangCooper1970}.

The second term in the brackets in Equation~\eqref{eq:Kompaneets_sph} accounts for photons cooling adiabatically as the ejecta expands (see Equation~\eqref{eq:varphi}). It can be shown using Equation~\eqref{eq:Kompaneets_sph} that when sources are absent, the average photon energy in zone $z$, ${\bar \epsilon}_z$, decreases due to adiabatic cooling as $d\log {\bar \epsilon}_z/d\log R = -\varphi_z$. However, the cooling of the photon field should cease when the radiation decouples from the plasma \citep{Peer2008, Beloborodov2011}. To account for this, the KRA shuts off adiabatic cooling above a specific radius, $R_{\rm cool}$, evaluated such that the total adiabatic cooling in the KRA equals the total cooling the photon field would experience in reality. The value of $R_{\rm cool}$ depends weakly on $\varphi$, but is roughly equal to $R_{\rm cool} \gtrsim R_{\rm ph}/1.2$.

\subsection{Source terms in \Kd}\label{subsub:source_terms}
In \Komrad, the number of photons in the RMS zone was assumed to be constant, i.e. $dN_{\rm out}/dR = dN_{\rm in}/dR$. From the discussion in Appendix~\ref{App:shock_broadening}, this implies an implicit assumption of ${\mathcal N}_{\rm sc} \ll \tau/2$. For \Kd, this assumption is relaxed. 

The outgoing number of photons is assumed to be given by Equation~\eqref{eq:dN_out_dr} (the expression before the last approximation). Since the probability to be advected from the shock zone into the downstream zone is assumed to be independent of the photon energy, the energy distribution of the escaping photons is the same as that of the photon distribution in the shock zone. The outgoing source term is thus
\begin{equation}\label{eq:s_out}
    s_{\rm out} = \frac{dN_{\rm out}}{dR} \frac{N_{r,\epsilon}}{N_r}  = \frac{N_{r,\epsilon}}{  c t_{{\rm sc}, r} \, {\mathcal N}_{\rm sc}} .
\end{equation}

To get a more accurate description of the incoming photon number, we do not assume that $\zeta_r$ and $\Delta \tau_r$ vary slowly with $R$ as in Equation~\eqref{eq:dN_r_dr}. The more general expression for the instantaneous change of photon number in the RMS is
\begin{equation}\label{eq:dN_dr_K}
    \frac{dN_r}{dR} = \frac{2}{R}N_r + \frac{4 \pi R^2}{\sigma_{\rm T} } \frac{d(\zeta \Delta \tau_r)}{dR}.
\end{equation}
\noindent In this paper, the evolution of $\zeta \Delta \tau_r$ with $R$ is deduced from the hydro simulation, assuming $\zeta = \zeta_u$. 

To assure the correct photon number in the RMS zone, the incoming photon number is given by
\begin{equation}\label{eq:dN_in_dr_K}
    \frac{d N_{\rm in}}{dR} = \frac{d N_r}{dR} + \frac{d N_{\rm out}}{dR},
\end{equation}
\noindent where $d N_r/dR$ is given by Equation~\eqref{eq:dN_dr_K} and $d N_{\rm out}/dR = N_r/{\mathcal N}_{\rm sc}/ c t_{{\rm sc},r}$.

The incoming photon number as given in Equation~\eqref{eq:dN_in_dr_K} is not necessarily positive. If the term is negative, this corresponds to the RMS width shrinking, and photons being ``advected'' into the upstream from the shock. \Kd\ allows for this by using the source term
\begin{equation}\label{eq:s_in}
    s_{\rm in} = \frac{dN_{\rm in}}{dR} \frac{N_{(u/r),\epsilon}}{N_{(u/r)}},
\end{equation}
\noindent where the subscript $(u/r)$ implies upstream zone quantities if $dN_{\rm in}/dR > 0$ and RMS zone quantities if $dN_{\rm in}/dR < 0$. The source terms used in Equation~\eqref{eq:Kompaneets_sph} are then
\begin{equation}\label{eq:s_r}
    s_r = s_{\rm in} - s_{\rm out}
\end{equation}
\noindent for the RMS zone(s) and 
\begin{equation}\label{eq:s_ds}
    s_{\ast} =  s^{\rm rs}_{\rm out} + s^{\rm fs}_{\rm out}
\end{equation}
\noindent for the common downstream zone. In the case of a single shock, the source term for the downstream will only contain one term.

That photons are allowed to flow from the RMS to the upstream means that the assumption of a thermal upstream photon distribution is not completely valid. However, in practice, this happens very rarely in the simulation and in this paper, we ignore this effect. 

To avoid injecting too many photons, the source term linking the upstream zone and the RMS zone is set to zero once the upstream has been depleted. The upstream is said to have been depleted once the condition
\begin{equation}\label{eq:upstream_depleted}
    \int \frac{dN_{\rm in}}{dR} \, dR \geq N_{u,{\rm tot}}
\end{equation}
\noindent is satisfied. After this time, the RMS zone still advects photons into the downstream zone according to Equation~\eqref{eq:s_out}. It also continues to evolve according to the Kompaneets equation.

\subsection{Energy conservation across the RMS zone}\label{subsec:energy_conservation}

The shock jump conditions dictate what the average energy of the photon distribution in the immediate downstream, ${\bar \epsilon}_d$, must be depending on the upstream conditions (see Equation~\eqref{eq:eps_d}). To ensure energy conservation in the KRA model, the photon distribution that is advected into the downstream zone should have an average energy that equals ${\bar \epsilon}_d$. According to Equation~\eqref{eq:s_out}, this implies that ${\bar \epsilon}_r = {\bar \epsilon}_d$ at all times, where ${\bar \epsilon}_r$ is the average energy in the RMS zone. In principle, this is achieved by choosing the appropriate value for $y_r$ as explained in Section \ref{subsec:KRA_breif_summary}. 

The appropriate value of $y_r$ when the RMS is in steady state can be obtained from the empirical relation given in \citet{Bagi2025} (see discussion in Section~\ref{subsec:H_KRA_RMS_to_KRA_conversion}). However, at early times during the shock formation and at late times during the dynamical stage of the shock, the radiation in the RMS will not be in steady state. To ensure energy conservation even at these times, the value of $y_r$ must be continuously adapted.

The Compton $y$-parameter for the RMS zone is defined as $y_r = 4\theta_r{\mathcal N}_{\rm sc}$. The value of $y_r$ can thus be changed by changing either ${\mathcal N}_{\rm sc}$ or $\theta_r$. 
In this paper, we assume that the change in $y_r$ necessary to satisfy the shock jump conditions is entirely due to a corresponding change in $\theta_r$. The real picture is likely more complicated with a correlated change to both quantities. Indeed, an increase in the energy gain per scattering across an RMS implies a steeper velocity gradient (in units of photon mean free paths), which in turn must effect the average number of scatterings. Such nonlinear effects require detailed numerical simulations, which are outside the scope of the current paper.

The RMS zone is initiated with a thermal Wien distribution at temperature $\theta_r = {\bar \epsilon}_d/3$ \citep[see Appendix A in][for motivation]{SamuelssonRyde2023}. Thus, the average photon energy is initially what it should be according to the shock jump conditions. Therefore, it is enough to ensure that $\partial{\bar \epsilon}_r/\partial R = \partial{\bar \epsilon}_d/\partial R$ to ensure ${\bar \epsilon}_r = {\bar \epsilon}_d$ for all $R$. Using Equation~\eqref{eq:Kompaneets_sph}, together with the source terms in Equations~\eqref{eq:s_out} and \eqref{eq:s_in}, it can be shown that the average photon energy in the RMS zone varies with radius as

\begin{equation}\label{eq:deps_d_dr}
    \frac{\partial{\bar \epsilon}_r}{\partial R} = \frac{4{\bar \epsilon}_r}{ct_{r,{\rm sc}}}(\theta_r - \theta_{r,{\rm C}}) - \frac{\varphi_r {\bar \epsilon}_r}{R} + \frac{1}{N_r}\frac{dN_{\rm in}}{dR}({\bar \epsilon}_{(u/r)} - {\bar \epsilon}_r),
\end{equation}

\noindent where ${\bar \epsilon}_{(u/r)}$ equals ${\bar \epsilon}_u$ if $dN_{\rm in}/dR > 0$ and ${\bar \epsilon}_r$ if $dN_{\rm in}/dR < 0$, and $\theta_{r, {\rm C}}$ is the Compton temperature in the RMS zone. Solving the above equation for $\theta_r$ and inserting $\partial{\bar \epsilon}_r/\partial R = \partial{\bar \epsilon}_d/\partial R$ gives the desired behaviour. We have ensured that the prescription works by comparing the input ${\bar \epsilon}_d(R)$ with ${\bar \epsilon}_r(R)$ as obtained in \Kd\ and the agreement is excellent.

For most of the simulation, $\theta_r$ is obtained from Equation~\eqref{eq:deps_d_dr} as explained above. However, at the final stages when the upstream has been depleted and the RMS zone is being drained of its remaining photons, $\theta_r$ is set equal to the Compton temperature in the RMS zone.

\subsection{Parameters for the dynamical KRA}\label{subsec:KRA_parameters}
Running a KRA simulation as described in this section requires input of the following parameters:\\ 
For the upstream zone(s),
\begin{itemize}
    \item $\theta_{u, {\rm  K}}(R)$, the temperature.
    \item $N_{u, {\rm tot}}$, the total photon number.
\end{itemize}
For the RMS zone(s),
\begin{itemize}
    \item ${\mathcal N}_{\rm sc}(R)$, the steady state number of scatterings.
    \item $\varphi_r(R)$, the adiabatic cooling term.
    \item $t_{r,{\rm sc}}(R)$, the observed scattering time. 
    \item ${\bar \epsilon}_d(R)$, the average photon energy in the immediate downstream.
\end{itemize}
For the downstream zone,
\begin{itemize}
    \item $\varphi_{\ast}(R)$, the adiabatic cooling term. 
    \item $t_{\ast,{\rm sc}}(R)$, the observed scattering time.
\end{itemize}
In addition to the parameters above, one also needs to know $\frac{d(\zeta \Delta \tau_r)}{dR} (R)$. However, this value can be obtained from the other parameters and is thus not a free parameter.

The above list results in eight parameters in the single shock case and fourteen parameters in the reverse plus forward shock case. Furthermore, most of these are functions of $R$, meaning that we need to now how they evolve with radius. However, it is important to note that these parameters may be coupled in non-trivial ways. Indeed, the hydrodynamical simulation in this paper only had eight free parameters but fixed all sixteen input parameters as a function of radius for the KRA.

Compared to its previous version, it is clear that the dynamical version of the KRA is much more complex. However, the lower complexity of the previous version can be recovered with the following simplifying assumption. Under the assumption of $\rho \propto R^{-2}$, one gets $\varphi_r = \varphi_{\ast} = 2/3$. Assuming that the shock(s) dissipates all of its energy during a doubling of the radius, $N_{u,{\rm tot}}$ is not a free parameter. Lastly, it was previously assumed that $t_{r, {\rm sc}}(R) = t_{\ast, {\rm sc}}(R)$ and that $\Gamma$ was constant. Under these assumptions, the KRA with all its new modifications can be run, and the subphotospheric evolution of the photon distribution obtained, knowing only the initial values of $\theta_{u,{\rm K}}$, ${\mathcal N}_{\rm sc}$, ${\bar \epsilon}_d$, and $\tau$, which is the same amount of free parameters as before.

\section{Obtaining the KRA input from the hydro output}\label{App:hydro_output_to_KRA_input}

\subsection{Converting between the RMS and the KRA parameters}\label{subsec:H_KRA_RMS_to_KRA_conversion}
As explained in Section~\ref{subsec:KRA_breif_summary}, a steady state RMS zone spectrum in the KRA is characterised by three parameters: the comoving upstream temperature $\theta_{u,{\rm K}}$, the effective temperature in the RMS zone, ${\tilde \theta}_r$, and the Compton $y$-parameter of the RMS zone, ${\tilde y}_r$. Here, the tilde indicates steady state. The number of scatterings in steady state is then obtained as ${\mathcal N}_{\rm sc} = {\tilde y}_r/4{\tilde \theta}_r$. 

The upstream temperature $\theta_{u,{\rm K}}$ is larger than $\theta_u$ by a factor of $(\rho_d/\rho_u)^{1/3}$ to account for the adiabatic compression that occurs across a real RMS \citep{BlandfordPayneII1981}. Thus, it is given by
\begin{equation}\label{eq:theta_u_K}
    \theta_{u,{\rm K}} = \left(\frac{\rho_d}{\rho_u}\right)^{1/3} \theta_u.
\end{equation}
\noindent The compression factor above is $\sim~$two for the relevant parameter space.

In \citet{Samuelsson2022}, the effective temperature ${\tilde \theta}_r$ was found by equating $\epsilon_{\rm max}^{\rm RMS} = \epsilon_{\rm max}^{\rm KRA}$. Via analytical arguments and an empirical study, it was determined that 
\begin{equation}\label{eq:theta_r}
    4{\tilde \theta}_r = \frac{(\gamma_u\beta_u)^2\ln({\bar \epsilon}_d/{\bar \epsilon}_u)}{\xi},
\end{equation}
\noindent and \citet{Samuelsson2022} found that $\xi \approx 55$ gives good agreement across a wide range of initial parameters.

The parameter ${\tilde y}_r$ is a measure of the energy increase of a typical photon within the RMS zone. It is obtained by requiring that the average energy of the photons that are advected into the downstream zone in the KRA equals the average energy immediately downstream of the shock in the hydrodynamical simulation, i.e. ${\bar \epsilon}_r = {\bar \epsilon}_d$.
In the KRA model, the value of ${\tilde y}_r$ corresponding to a specific average photon energy $\bar{\epsilon}_r$ is not known a priori.
In principle, a simulation of the Kompaneets equation can be run in steady state iteratively until the unique value of ${\tilde y}_r$ that corresponds to ${\bar \epsilon}_r$ is found. To save computational time, here, we employed the empirical relation between ${\tilde y}_r$ and ${\bar \epsilon}_r$ that is given in \citet{Bagi2025}. The relation is accurate to within 5\% in the relevant parameter region.

\subsection{Scattering time, adiabatic cooling term, and evolution of \texorpdfstring{$\zeta \Delta \tau_r$}{TEXT}}\label{subsec:H_KRA_scattering_adcool}
As mentioned in Section~\ref{subsec:H_obtaining_properties}, the values for the common downstream zone density and bulk Lorentz factor were taken as the mass averaged value of each quantity between the two shock fronts. With $\rho_\ast$ and $\Gamma_\ast$ obtained, $\varphi_\ast$ and $t_{\ast,{\rm sc}}$ were calculated using Equations~\eqref{eq:varphi} and \eqref{eq:t_sc}, respectively.

The scattering time in the RMS zone(s) was slightly more complicated, since the plasma density and bulk Lorentz factor changes significantly across the RMS transition region. In this paper, we used
\begin{equation}\label{eq:t_sc_r}
    t_{r,{\rm sc}}^{-1} = \frac{t_{d,{\rm sc}}^{-1} + t_{u,{\rm sc}}^{-1}}{2},
\end{equation}
\noindent that is, we set the RMS zone scattering frequency to be the average of the upstream and immediate downstream scattering frequencies. In the adiabatic cooling term, we used the immediate downstream value for the density.

To solve Equation~\eqref{eq:dN_dr_K}, we needed $d(\zeta \Delta \tau_r)/dR$. For the photon-to-proton ratio, we used the value from Equation~\eqref{eq:zeta} evaluated in the upstream region. The optical depth across the shock was evaluated in the non-relativistic limit as $\Delta \tau_r = \sqrt{{\mathcal N}_{\rm sc}} = \sqrt{{\tilde y}_r/4{\tilde \theta}_r}$.

\subsection{Initial photon distributions}\label{subsec:H_KRA_N_init}
In this paper, the downstream zone was not empty at the beginning of the simulation. This is because the initial plasma is rather hot, and the reverse and the forward shocks did not form at the same mass coordinate. Let us denote the mass coordinate where the reverse shock formed by $m_{\rm rs}$, and the mass coordinate where the forward shock formed by $m_{\rm fs}$, where $m_{\rm fs} > m_{\rm rs}$. The number of photons in the downstream at the start of the simulation was then given by
\begin{equation}\label{eq:N_ds_init}
    N_{{\ast}, {\rm init}}= \int_{m_{\rm rs}}^{m_{\rm fs}} \zeta(m)\frac{dm}{m_p}.
\end{equation}
\noindent These photons were put in a thermal Wien distribution at temperature $\theta_{\ast} = w_{\rm cd} m_p/4m_e\zeta_{\rm cd}$ (see Equation~\eqref{eq:theta_u}). Here, $w_{\rm cd}$ and $\zeta_{\rm cd}$ are the dimensionless enthalpy and the photon-to-proton ratio at the contact discontinuity, respectively.

As mentioned in Section \ref{subsec:energy_conservation}, the RMS zone(s) were initiated with a thermal Wien distribution at a temperature of ${\bar \epsilon}_d/3$. The number of photons in this initial distribution was given by (see Equation~\eqref{eq:N_r})
\begin{equation}\label{eq:N_r_init}
    N_{r, \rm init} = \frac{4 \pi R^2 \zeta \Delta \tau_r}{\sigma_{\rm T} },
\end{equation}
\noindent where the values of $R$, $\zeta$, and $\Delta \tau_r$ were taken to be the respective values at the radius where the shock forms. In the hydro simulation, the forward shock and the reverse shock did not form at the same time. The KRA simulation was set to start once the first of the two shocks had formed. The second RMS zone was included once the KRA simulation had reached the shock formation radius of the second RMS. 

The total number of photons that would pass through each shock was obtained by integrating the photon number in between the mass coordinate where the shock started ($m_{\rm rs}$ or $m_{\rm fs}$) and the mass coordinate where the shock would end ($m_{\rm ll}$ or $m_{\rm ul}$). The initial total photon number in the upstream was then given by
\begin{equation}\label{eq:N_u_tot}
    N_{u,{\rm tot}} = \int_{m_1}^{m_2} \zeta(m)\frac{dm}{m_p} - N_{r, {\rm init}},
\end{equation}
\noindent where the mass coordinate limits $m_1$ and $m_2$ equal $m_{\rm ll}$ ($m_{\rm fs}$) and $m_{\rm rs}$ ($m_{\rm ul}$) for the reverse shock upstream (forward shock upstream), respectively. 

\end{appendix}

\end{document}